\documentclass[a4paper,UKenglish,autoref,thm-restate]{lipics-v2021}

\hideLIPIcs
\nolinenumbers

\usepackage[utf8]{inputenc}
\usepackage[T1]{fontenc}

\usepackage{xparse}
\usepackage{xpatch}
\usepackage{tokcycle}

\usepackage[obeyclassoptions,mode=tex]{standalone}

\usepackage{amsmath}
\usepackage{amsthm}
\usepackage{thmtools}
\usepackage{upgreek}
\usepackage{amssymb}
\usepackage{stmaryrd}

\usepackage{hyperref}
\usepackage[capitalise,noabbrev,nameinlink]{cleveref}
\usepackage[electronic,hyperref,xcolor,cleveref]{knowledge}
\knowledgeconfigure{notion}

\usepackage{booktabs}
\usepackage{varwidth}

\usepackage{bussproofs}

\usepackage{tikz}
\usetikzlibrary{backgrounds}
\usetikzlibrary{shapes.geometric}
\usetikzlibrary{positioning}
\usetikzlibrary{automata}
\usetikzlibrary{tikzmark}
\usetikzlibrary{patterns}
\usetikzlibrary{arrows}
\usepackage{tikz-cd}
\tikzset{every state/.style={minimum size=1pt}}

\definecolor{Prune}{RGB}{99,0,60}
\definecolor{A1}{HTML}{000000}
\definecolor{B1}{RGB}{49,62,72}
\definecolor{C1}{RGB}{124,135,143}
\definecolor{D1}{RGB}{213,218,223}
\definecolor{A2}{RGB}{198,11,70}
\definecolor{B2}{RGB}{237,20,91}
\definecolor{C2}{RGB}{238,52,35}
\definecolor{D2}{RGB}{243,115,32}
\definecolor{A3}{RGB}{124,42,144}
\definecolor{B3}{RGB}{125,106,175}
\definecolor{C3}{RGB}{198,103,29}
\definecolor{D3}{RGB}{254,188,24}
\definecolor{A4}{RGB}{0,78,125}
\definecolor{B4}{RGB}{14,135,201}
\definecolor{C4}{RGB}{0,148,181}
\definecolor{D4}{RGB}{70,195,210}
\definecolor{A5}{RGB}{0,128,122}
\definecolor{B5}{RGB}{64,183,105}
\definecolor{C5}{RGB}{140,198,62}
\definecolor{D5}{RGB}{213,223,61}

\colorlet{context}{B4}
\colorlet{rightlabel}{A2}
\colorlet{leftlabel}{A3}
\colorlet{embedding}{A5}

\foreach \name in {A,B,C,D} {
    \foreach \hue in {1,2,3,4,5} {
        \foreach \shade/\intensity in {hint/20,bg/50} {
            \xglobal\colorlet{\name\hue\shade}{\name\hue!\intensity!white}
        }
    }
}

\usepackage{crossreftools}
\knowledgestyle{intro notion}{color={A5}, emphasize}
\knowledgestyle{notion}{color={A4}}
\knowledgeconfigure{anchor point color={A2},
                    anchor point shape=corner}
\knowledgestyle{intro unknown}{color={D3}, emphasize}
\knowledgestyle{intro unknown cont}{color={C3}, emphasize}
\knowledgestyle{kl unknown}{color={D2}}
\knowledgestyle{kl unknown cont}{color={C2}}

\hypersetup{
    colorlinks=true,
    anchorcolor=A2,
    citecolor=A4,
    linkcolor=A4,
    urlcolor=A3,
    filecolor=A3,
    runcolor=D2,
    menucolor=D2,
}

\NewDocumentCommand{\klscope}{ o m }{
    \withkl{\kl[#1]}{#2}
}

\makeatletter
\newcommand\mathgr[1]{\tokcycle
  {\addcytoks{##1}}
  {\processtoks{##1}}
  {\ifcsname up\expandafter\@gobble\string##1\endcsname
   \addcytoks[1]{\csname up\expandafter\@gobble\string##1\endcsname}%
    \else\addcytoks{##1}\fi}
  {\addcytoks{##1}}{#1}%
  \expandafter\mathrm\expandafter{\the\cytoks}%
}
\makeatother

\NewDocumentCommand{\citek}{ o m }{
    \IfNoValueTF{#1}{
        \cite{m}
    }{
        \cite[{\kl(#1)[#2]}]{#1}
    }
}

\NewDocumentEnvironment{proofof}{ m }{
    \ifcsname #1\endcsname
        \def\isInsideRestatedTheorem{1}
        \csname #1\endcsname*
    \fi
    \begin{proof}[Proof of {\cref{#1}} page {\pageref{#1}}]
        \phantomsection
        \label{#1:proof}
}{

        \noindent\hyperref[#1]{$\triangleright$ Go back to \cref{#1} on page \pageref{#1}}
    \end{proof}
}

\NewDocumentCommand{\proofref}{ m }{
    \IfRefUndefinedExpandable{#1:proof}{}{
        \ifdefined\isInsideRestatedTheorem
        \else
        \hfill\hyperref[#1:proof]{\textsf{(Go to proof p.\pageref{#1:proof})}}
        \fi
    }
}
\NewDocumentCommand{\defined}{}{\mathrel{:=}}
\NewDocumentCommand{\set}{ m }{\left\{#1\right\}}
\NewDocumentCommand{\setof}{ m O{\mid} m}{\{ #1 #2 #3 \}}
\NewDocumentCommand{\seqof}{ m m }{\left(#1\right)_{#2}}

\NewDocumentCommand{\image}{ m }{\operatorname{Img}(#1)}

\NewDocumentCommand{\Bool}{}{\mathbb{B}}

\NewDocumentCommand{\Nat}{}{\mathbb{N}}

\NewDocumentCommand{\MSO}{}{\mathsf{MSO}}
\NewDocumentCommand{\FO}{}{\mathsf{FO}}

\NewDocumentCommand{\vertices}{}{\mathsf{V}}
\NewDocumentCommand{\edges}{}{\mathsf{E}}
\NewDocumentCommand{\isubleq}{}{\subseteq_i}
\NewDocumentCommand{\Cls}{ O{C} }{\mathcal{#1}}
\NewDocumentCommand{\Label}{ m }{\mathsf{Label}_{#1}}

\NewDocumentCommand{\vlbl}{}{\mathsf{l}}

\undef\Alph
\NewDocumentCommand{\Alph}{O{\Sigma}}{#1}

\NewDocumentCommand{\AGlue}{O{Gl}}{\mathsf{#1}}

\NewDocumentCommand{\wleq}{O{\leq}}{%
    #1^\star%
}
\NewDocumentCommand{\gwleq}{O{\leq} O{\AGlue}}{%
    #1_{#2}^\star%
}

\NewDocumentCommand{\bp}{}{\mathrel{%
    \tikz[line cap=round, line join=round]{%
    \draw
      (0ex, 1.3ex) -- (0ex, 0.6ex) -- (-0.5ex, 0.6ex) -- (-0.5ex, 0ex);
    \draw
      (0ex, 1.3ex) -- (0ex, 0.6ex) -- (0.5ex, 0.6ex) -- (0.5ex, 0ex);
      }%
  }%
}

\NewDocumentCommand{\ep}{}{\mathop{%
    \tikz[line cap=round, line join=round]{%
    \draw
      (0ex, 1.3ex) -- (0ex, 0.7ex) -- (-0.5ex, 0.7ex) -- (-0.5ex, 0ex);
    \draw
      (0ex, 1.3ex) -- (0ex, 0.7ex) -- (0.5ex, 0.7ex) -- (0.5ex, 0ex);
    \draw
      (0ex, 1.3ex) -- (0ex, 0ex);
    \draw
      (-0.5ex, 0.5ex) -- (0.5ex, 0.5ex);
      }%
  }%
}

\NewDocumentCommand{\ftype}{ O{F} m }{ \operatorname{tp}_{#1}(#2)}
\NewDocumentCommand{\frst}{ O{} m }{\mathcal{F}_{#1}(#2)}

\NewDocumentCommand{\jideal}{ m }{\kl[\jideal]{(#1)_{\mathfrak{J}}}}
\NewDocumentCommand{\lideal}{ m }{\kl[\lideal]{(#1)_{\mathfrak{L}}}}
\NewDocumentCommand{\rideal}{ m }{\kl[\rideal]{(#1)_{\mathfrak{R}}}}
\knowledge{\jideal}{notion}
\knowledge{\lideal}{notion}
\knowledge{\rideal}{notion}
\NewDocumentCommand{\jleq}{}{ \mathrel{\kl[\jleq]{\leq_{\mathfrak{J}}}}}
\NewDocumentCommand{\jequiv}{}{\mathrel{\kl[\jequiv]{\equiv_{\mathfrak{J}}}}}
\knowledge{\jleq}{notion}
\knowledge{\jequiv}{notion}

\NewDocumentCommand{\dalileq}{}{\mathrel{\preccurlyeq}}

\NewDocumentCommand{\aMgraph}{ O{G} }{\mathfrak{#1}}

\NewDocumentCommand{\mcast}{ m }{ \withkl{\kl[\mcast]}{\cmdkl{\lfloor} #1 \cmdkl{\rfloor}}}
\knowledge{\mcast}{notion}

\NewDocumentCommand{\unit}{ O{M}}{\mathbf{1}_{#1}}

\NewDocumentCommand{\RegMG}{ m }{\mathsf{MGraph}_{#1}}

\NewDocumentCommand{\aTreeModel}{ O{T} }{\mathfrak{#1}}
\NewDocumentCommand{\tmeval}{}{\mathsf{treeEval}}

\NewDocumentCommand{\dwclosure}{ O{\leq} m }{\mathop{{\downarrow}_{#1}}#2}

\NewDocumentCommand{\msoIm}{}{\mathop{\mathsf{Im}}}

\NewDocumentCommand{\layered}{}{ \mathop{\kl[\layered]{\mathsf{lay}}}}
\knowledge{\layered}{notion}

\NewDocumentCommand{\tmleq}{}{\mathrel{\kl[\gapleq]{\leq_{\mathsf{model}}}}}
\knowledge{\tmleq}{notion}

\NewDocumentCommand{\gapleq}{}{\mathrel{\kl[\gapleq]{\leq_{\mathsf{gap}}}}}
\knowledge{\gapleq}{notion}

\NewDocumentCommand{\mtogap}{}{\mathop{\kl[\mtogap]{\mathsf{mToGap}}}}
\knowledge{\mtogap}{notion}
\knowledge{notion}
 | graphs

\knowledge{notion}
 | embedding@graph
 | embeddings@graph
\knowledge{notion}
 | induced extension
\knowledge{notion}
 | induced subgraph
 | induced subgraphs
 | induced subgraph relation
\knowledge{notion}
 | hereditary@class
 | hereditary class
 | hereditary classes
\knowledge{notion}
 | hereditary closure
\knowledge{notion}
 | good@sequence
 | good sequence
\knowledge{notion}
 | bad@sequence
 | bad sequence
\knowledge{notion}
 | well-founded
\knowledge{notion}
 | well-quasi-order
 | well-quasi-ordered
 | well-quasi-orderings
 | well-quasi-ordering
\knowledge{notion}
 | Hoare quasi-ordering relation
\knowledge{notion}
 | labelled@class
 | labelled graph
\knowledge{notion}
 | colored@class
\knowledge{notion}
 | embedding@labelled
\knowledge{notion}
 | embedding@colored
\knowledge{notion}
 | $k$-well-quasi-ordered
 | $2k$-well-quasi-ordered
 | $(k+1)$-well-quasi-ordered
 | $2$-well-quasi-ordered
 | $2$-well-quasi-ordering
 | $2$-well-quasi-orderings
 | $k$-wqo
 | $2$-wqo
\knowledge{notion}
 | labelled-well-quasi-ordering
 | labelled-well-quasi-orderings
 | labelled-well-quasi-ordered
 | labelled-wqo
\knowledge{notion}
 | $\infty$-well-quasi-ordered
 | $\infty$-well-quasi-ordering
 | $\infty$-well-quasi-orderings
 | $\infty$-wqo
\knowledge{notion}
 | $\MSO $-interpretation
 | $\MSO $-interpretations
\knowledge{notion}
 | monadic second-order logic
\knowledge{notion}
 | bounded linear clique-width
\knowledge{notion}
 | bounded clique-width
\knowledge{notion}
 | $\AGlue $-embedding
\knowledge{notion}
 | subword embedding
 | subword ordering
\knowledge{notion}
 | factorization forest
 | factorization forests
\knowledge{notion}
 | type of a leaf
 | type
\knowledge{notion}
 | inconsistency graph
\knowledge{notion}
 | inconsistency relation
\knowledge{notion}
 | consistency relation
\knowledge{notion}
 | bad idempotent paths
 | bad idempotent path
\knowledge{notion}
 | infix

\knowledge{notion}
 | good forest path
 | good forest paths
\knowledge{notion}
 | bad forest path
 | bad forest paths

\knowledge{notion}
 | monoid
\knowledge{notion}
 | morphism
\knowledge{notion}
 | idempotent
\knowledge{notion}
 | monoid interpretation
 | monoid interpretations

\knowledge{notion}
 | bilateral ideal
 | bilateral ideals
\knowledge{notion}
 | left ideal
 | left ideals
\knowledge{notion}
 | right ideal
 | right ideals

\knowledge{notion}
 | downwards closed
\knowledge{notion}
 | downwards closure

\knowledge{notion}
 | monoid-labelled graphs
 | monoid-labelled graph

\knowledge{notion}
 | $\MSO $-transductions
 | $\MSO $-transduction

\knowledge{notion}
 | monoid downcasting
 | monoid downcastings

\knowledge{notion}
 | edge selector
 | edge selectors

\knowledge{notion}
 | binary product
\knowledge{notion}
 | idempotent product
 | idempotent products

\knowledge{notion}
 | regular monoid-labelled graphs
 | regular monoid-labelled

\knowledge{notion}
 | gap embedding relation
 | gap embedding

\knowledge{notion}
 | order reflection

\knowledge{notion}
 | totally ordered
 | composition ordering
 | totally ordered monoid
 | totally ordered monoids
 | finite power property

\knowledge{notion}
 | evaluates@mlg
 | evaluate@mlg
 | evaluating@mlg
 | evaluation@mlg
 | evaluations@mlg

\knowledge{notion}
 | image@interpretation

\knowledge{notion}
 | Kruskal embeddings
 | Kruskal embedding

\knowledge{notion}
 | tree-model
 | tree-models
 | tree model
 | tree models

\knowledge{notion}
 | tree-model embedding
 | tree model embedding
 | tree-model embedding relation
 | embedding@tree-model

\knowledge{notion}
 | layered interpretation

\knowledge{notion}
 | split
 | splitting
 | 
 
\crefname{conjecture}{Conjecture}{Conjectures}
\Crefname{conjecture}{Conjecture}{Conjectures}
\crefname{fact}{Fact}{Facts}
\Crefname{fact}{Fact}{Facts}
\crefname{algocf}{Algorithm}{Algorithms}
\Crefname{algocf}{Algorithm}{Algorithms}

\begin{document}
\title{Labelled Well Quasi Ordered Classes of Bounded Linear Clique-Width}

\newcommand{\acknowledge}{}

\newcommand{\makeabstract}{
    \begin{abstract}

        We are interested in characterizing which classes of finite graphs are
        well-quasi-ordered by the induced subgraph relation. To that end, we
        devise an algorithm to decide whether a class of finite graphs
        well-quasi-ordered by the induced subgraph relation when the vertices
        are labelled using a finite set. In this process, we answer positively
        to a conjecture of Pouzet, under the extra assumption that the class is
        of bounded linear clique-width. As a byproduct of our approach, we
        obtain a new proof of an earlier result from Daliagault, Rao, and
        Thomass\'e, by uncovering a connection between well-quasi-orderings on
        graphs and the gap embedding relation of Dershowitz and Tzameret.

    \end{abstract}
}

\author{Aliaume Lopez}%
       {University of Warsaw}%
       {ad.lopez@uw.edu.pl}%
       {https://orcid.org/0000-0002-1825-0097}%
       {}

\authorrunning{A. Lopez}

\titlerunning{Well Quasi Orders and Linear Clique-Width}
\date{\today}

\Copyright{Aliaume LOPEZ}

\category{} %
\relatedversion{} %


\EventEditors{John Q. Open and Joan R. Access}
\EventNoEds{2}
\EventLongTitle{42nd Conference on Very Important Topics (CVIT 2016)}
\EventShortTitle{CVIT 2016}
\EventAcronym{CVIT}
\EventYear{2016}
\EventDate{December 24--27, 2016}
\EventLocation{Little Whinging, United Kingdom}
\EventLogo{}
\SeriesVolume{42}
\ArticleNo{23}
\maketitle
\makeabstract
\acknowledge

\section{Introduction}
\label{introduction:sec}

A cornerstone result of structural graph theory is the Graph Minor Theorem of Robertson
and Seymour \cite{ROBSEY04}, which states that the class of all graphs is
well-quasi-ordered by the graph minor relation. That is, given any infinite
sequence of graphs, there exists an increasing pair of graphs in the sequence
such that the first is a minor of the second. This result has profound
implications in graph theory, and algorithmic consequences such as the
polynomial-time solvability of whether a graph can be embedded into a given
surface \cite{ROBSEY04}.

More generally, as generic combinatorial objects, \kl{well-quasi-orderings}
have been used in various areas of computer science, one particular application
being to the proof of termination for algorithms, and more specifically in the
verification of \emph{Well-Structured Transition Systems}
\cite{ABDU96,ABDU98,FINSCH01}. One of the appeal of this theory is the
existence of rich algebra of constructions that preserve well-quasi-orderings,
such as finite sums \cite{SCSC12}, finite products \cite[Dickson's
Lemma]{SCSC12}, finite words \cite{HIG52}, and finite trees \cite{KRU72}, which
makes them a powerful tool to reason about systems where the state space is
infinite. In this setting, the Graph Minor Theorem can be seen as a particular
instance of a construction that preserves well-quasi-orderings, namely
considering finite graphs ordered using the \emph{minor relation}.

Unfortunately, the analogues of the Graph Minor Theorem for other types of
embeddings between graphs, such as the induced subgraph relation, the situation
is more complex: some classes of graphs are well-quasi-ordered by the induced
subgraph relation (such as finite paths) while others are not (such as finite
cycles). It is believed that a better-behaved notion is to ask whether a class
of graph is well-quasi-ordered by the induced subgraph relation when the
vertices are labelled, i.e., considering \emph{colored} graphs. For instance,
neither cycles and paths are well-quasi-ordered by the induced subgraph
relation when endowed with two labels. This intuition is consistent with the
interpretation of the Graph Minor Theorem as a \emph{constructor} for new
well-quasi-ordered sets, which even suggests that the correct notion would be
to color graphs using a well-quasi-ordered set of labels. Furthermore, it is
common to restrict our attention to \emph{hereditary} classes of graphs, i.e.,
closed under induced subgraphs, that are adapted to the induced subgraph
ordering.

In this context, the first Pouzet conjecture, restated here as
\cref{pouzet1:conj}, states that hereditary classes of graphs are
well-quasi-ordered by the induced subgraph relation when extended with $2$
labels if and only if they are well-quasi-ordered by this relation when
extended with \emph{any} finite number of labels. This conjecture essentially
states that while being well-quasi-ordered and being well-quasi-ordered with $k
\geq 2$ labels are different notions (finite paths vs finite cycles), the
precise value of $k \geq 2$ does not matter. A variant of this conjecture,
stated in \cref{pouzet2:conj}, is moving from finite sets of labels to being
labelled by a well-quasi-ordered set of labels. We attribute the second version
to Pouzet as well, but it is not explicitly stated in \cite{POUZ72}.

\begin{conjecture}[Pouzet 1 {\cite{POUZ72}}]
    \label{pouzet1:conj}
    A hereditary class of graphs is well-quasi-ordered by the induced subgraph relation with 
    finitely many labels if and only if it is well-quasi-ordered by the induced
    subgraph relation when the vertices are labelled with at most $2$ labels.
\end{conjecture}

\begin{conjecture}[Pouzet 2]
    \label{pouzet2:conj}
    A class of graphs is well-quasi-ordered by the induced subgraph relation with
    vertices labelled using a well-quasi-ordered set of labels if and only if it
    is well-quasi-ordered by the induced subgraph relation when the vertices are
    labelled with at most $k$ labels, for any choice of $k \in \Nat$.
\end{conjecture}

In 2010, Daligault Rao and Thomass\'e characterized which monoids of so-called
\emph{relabel functions} give rise to classes of graphs that are
well-quasi-ordered by the induced subgraph relation with two labels
\cite[Theorem 3]{DRT10}. For such classes constructed using \emph{relabel
functions}, both Pouzet conjectures were confirmed \cite[page 4]{DRT10}. The
choice of representing graphs using relabel functions is motivated by the fact
that they provide an (asymptotic) equivalent characterization of classes of
graphs of bounded clique-width called $\mathsf{NLC}$-width
\cite{WANKE94,COJOGR93}. Because of this connection, it was also conjectured
that hereditary $2$-well-quasi-ordered classes of graphs are all of bounded
clique width \cite[Conjecture 5, p.\ 13]{DRT10}. However, this would not
provide a positive answer to either of Pouzet's conjectures, as the
characterization of \cite{DRT10} cannot be applied on a single class of bounded
clique-width. In the meantime, some of their other conjectures already have
been disproved: for instance, being ``simply'' well-quasi-ordered does not
imply having bounded clique width \cite{LRZ16}. Other attempts at
characterizing classes of graphs that are well-quasi-ordered for the induced
subgraph relation with labels have been made \cite{DLP17,POZA22},  but these
results are often leveraging deep understanding of the structure of the classes
of graphs under consideration, and are not easily generalizable.

\subparagraph*{Contributions} Instead of trying to characterize classes of
graphs that are well-quasi-ordered by the induced subgraph relation in full
generality, we provide an \emph{algorithm} that inputs a class of graphs of
bounded linear clique-width (given as an $\MSO$ interpretation) and decide
whether it is well-quasi-ordered by the induced subgraph relation when the
vertices are labelled using $k \geq 2$ labels.

Restricting our attention to classes of graphs obtained by interpreting an
$\MSO$ formula over finite words allows us to leverage classical tools from
automata theory, such as the factorization forest theorem of Simon
\cite{SIMO90}. Furthermore, our proof scheme implies a weak version of the
Pouzet conjecture: for any such class $\Cls$, there exists a computable number
$k$ of labels, such that $\Cls$ is well-quasi-ordered by the induced subgraph
relation when the vertices are labelled with at most $k$ labels if and only if
$\Cls$ is well-quasi-ordered using with labels taken from a well-quasi-ordered
set. This implies \cref{pouzet2:conj}, and acts as a parametrized version of
\cref{pouzet1:conj}, which is a significant step towards a full resolution of
both Pouzet conjectures. Finally, as a byproduct of our effective decision
procedure, we can answer \cref{pouzet2:conj} positively for all classes $\Cls$
of bounded linear clique-width.

We believe that our approach follows a promising direction,
as it avoids the use of complex \emph{minimal bad sequence arguments}
\cite{NASH65}, and instead relies on suitable combination of classical
constructions preserving well-quasi-orderings (finite words \cite{HIG52},
finite products \cite[Dickson's Lemma]{SCSC12}). To further justify this
statement, we also propose an automata theoretic approach to the earlier result
of \cite{DRT10}, based on well studied classes of monoids that recognize
languages of finite words having the so-called \emph{finite power property}
\cite{KIRS02,KUNC05}. This, combined with the usage of the \emph{gap embedding
relation} on trees \cite{DERSHOWITZ200380}, provides a self-contained
alternative proof of the result of \cite{DRT10}, that also avoids \emph{minimal
bad sequence arguments}.

\subparagraph{Outline} In \cref{preliminaries:sec}, we introduce the
definitions and notations regarding \kl{graphs}, \kl(graph){embeddings},
\kl{well-quasi-orderings} and \kl{$\MSO$-interpretations}. Then, in
\cref{bounded-lcw:sec}, we prove our main \cref{main-theorem:thm} and
its \cref{main-corollary:cor} that answers to \cref{pouzet2:conj} in
the case of classes of \kl{bounded linear clique-width}. Finally, in
\cref{totally-ordered-monoids:sec}, we give a \emph{minimal bad sequence
argument free} proof of the result of \cite{DRT10} in
\cref{gap-embedding-tree-model:thm}, and discuss in
\cref{good-specialisation:thm} the relationship between the results of
\cite{DRT10} and our \cref{main-theorem:thm}.
\section{Preliminaries}
\label{preliminaries:sec}

\AP In this paper, \intro{graphs} are all finite, simple and undirected unless
otherwise explicitly stated. We denote by $\vertices(G)$ the set of vertices of
a graph $G$, and by $\edges(G)$ the set of edges. A map $h \colon G \to H$
between two graphs $G$ and $H$ is an \intro(graph){embedding} if it is
injective and satisfies $\set{h(u),h(v)} \in \edges(H)$ if and only if
$\set{u,v} \in \edges(G)$, for all $(u,v) \in \vertices(G)$. We write $G
\isubleq H$ if there exists an embedding $h \colon G \to H$, and signifies that
$G$ is an \intro{induced subgraph} of $H$, or equivalently that $H$ is an
\intro{induced extension} of $G$. A class of graph is \intro(class){hereditary}
when it is closed under taking induced subgraphs. We denote by
$\dwclosure[\isubleq]{\Cls}$ the \intro{hereditary closure} of a class $\Cls$,
obtained by considering all \kl{induced subgraphs} of graphs in $\Cls$.

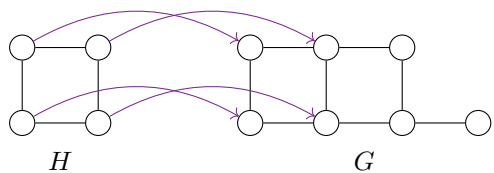
\begin{figure}[h]
    \centering
    \begin{tikzpicture}
        \node[circle,draw] (H1) at (0,0) {};
        \node[circle,draw] (H2) at (1,0) {};
        \node[circle,draw] (H3) at (1,1) {};
        \node[circle,draw] (H4) at (0,1) {};
        \draw (H1) -- (H2) -- (H3) -- (H4) -- (H1);
        \node at (0.5,-0.5) {$H$};
        \node[circle,draw] (G1) at (3,0) {};
        \node[circle,draw] (G2) at (4,0) {};
        \node[circle,draw] (G3) at (4,1) {};
        \node[circle,draw] (G4) at (3,1) {};
        \node[circle,draw] (G5) at (5,0) {};
        \node[circle,draw] (G6) at (5,1) {};
        \node[circle,draw] (G7) at (6,0) {};
        \draw (G1) -- (G2) -- (G3) -- (G4) -- (G1);
        \draw (G2) -- (G5) -- (G6) -- (G3);
        \draw (G5) -- (G7);
        \node at (4.5,-0.5) {$G$};
        \draw[->,A3] (H1) to[bend left=30] (G1);
        \draw[->,A3] (H2) to[bend left=30] (G2);
        \draw[->,A3] (H3) to[bend left=30] (G3);
        \draw[->,A3] (H4) to[bend left=30] (G4);
    \end{tikzpicture}
    \caption{
        The graph $H$ is an induced subgraph of $G$,
        the arrows between vertices in $H$ and vertices in $G$
        represent an \kl(graph){embedding} of $H$ into $G$.
    }
    \label{induced-subgraph:fig}
\end{figure}

\AP Let $(X, \leq)$ be a quasi-ordered set.
We use the notation 
$\dwclosure{X}$ for the \intro{downwards closure}
of a set $X$ with respect to the quasi-ordering $\leq$,
defined by $\dwclosure{X} \defined \setof{y \in X}{y \leq x}$ for all $x \in X$.
A set $X$ such that $\dwclosure{X} = X$ is \intro{downwards closed}.
Note that the \kl{hereditary closure} of a class $\Cls$ is the
\kl{downwards closure} of $\Cls$ for $\isubleq$.
A sequence $\seqof{x_i}{i \in
\Nat}$ is \intro(sequence){good} if there exists $i < j$ such that $x_i \leq
x_j$, and is \intro(sequence){bad} otherwise. A quasi-ordered set $(X, \leq)$
is \intro{well-quasi-ordered} if it does not contain an infinite \kl{bad
sequence}. We refer the reader to \cite{SCSC12} for a comprehensive survey on
the topic. 

\begin{example}
    \label{well-quasi-ordered:ex}
    The set $(\Nat, \leq)$ is \kl{well-quasi-ordered},
    but $(\Nat, =)$ is not \kl{well-quasi-ordered}, although both are \kl{well-founded}.
    In fact, a set $(X, =)$ is \kl{well-quasi-ordered} if and only if it is finite.
\end{example}

\AP We use cursive letters $\Cls[A]$, $\Cls[B]$, $\Cls[C]$ to denote classes of
graphs. Given a class $\Cls$ of graphs, and a set $X$, we denote
$\Label{X}(\Cls)$ the class of graphs equipped with a function $\vlbl \colon
\vertices(G) \to X$ such that $G \in \Cls$. We say that $\Label{X}(\Cls)$ is
the class of $X$-\intro(class){labelled} graphs of $\Cls$. 
Let $\Cls$ be a class of graphs, and $(X, \leq)$ be a quasi-ordered set. We
extend the notion of embedding to $X$-labelled graphs as follows: a map $h
\colon G \to H$ between two $X$-labelled graphs $G$ and $H$ is an
\intro(labelled){embedding} if it is an embedding of the underlying graphs, and satisfies
$\vlbl(h(u)) \leq \vlbl(u)$ for all $u \in \vertices(G)$.

\begin{figure}[h]
    \centering
    \begin{tikzpicture}
        \node[circle,draw,fill] (H1) at (0,0) {};
        \node[circle,draw] (H2) at (1,0) {};
        \node[circle,draw] (H3) at (1,1) {};
        \node[circle,draw] (H4) at (0,1) {};
        \draw (H1) -- (H2) -- (H3) -- (H4) -- (H1);
        \node at (0.5,-0.5) {$H$};
        \node[circle,draw] (G1) at (3,0) {};
        \node[circle,draw] (G2) at (4,0) {};
        \node[circle,draw] (G3) at (4,1) {};
        \node[circle,draw] (G4) at (3,1) {};
        \node[circle,draw] (G5) at (5,0) {};
        \node[circle,draw] (G6) at (5,1) {};
        \node[circle,draw,fill] (G7) at (6,0) {};
        \draw (G1) -- (G2) -- (G3) -- (G4) -- (G1);
        \draw (G2) -- (G5) -- (G6) -- (G3);
        \draw (G5) -- (G7);
        \node at (4.5,-0.5) {$G$};
    \end{tikzpicture}
    \caption{
        We represent graphs $G$ and $H$ using filled vertices to
        represent the presence of a label.
        Recall that $H$ is a subgraph of $G$ if we discard the labelling,
        as illustrated in \cref{induced-subgraph:fig}. In this case,
        the graph $H$ is not an \kl{induced subgraph} of the graph $G$
        because of the labelling of the vertices.
    }
\end{figure}
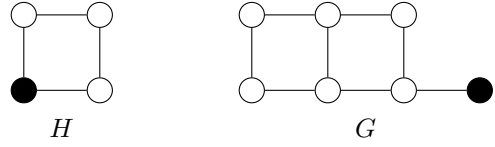

\AP We say that a class $\Cls$ of graphs is \intro{$k$-well-quasi-ordered} if
$\Label{X}(\Cls)$ is well-quasi-ordered for all finite sets $(X,=)$ of size at
most $k$. We say that $\Cls$ is \intro{$\infty$-well-quasi-ordered} if
$\Label{X}(\Cls)$ is well-quasi-ordered for all finite sets $(X,=)$. We say
that $\Cls$ is \intro{labelled-well-quasi-ordered} if $\Label{X}(\Cls)$ is
well-quasi-ordered for all \kl{well-quasi-ordered} sets $(X,\leq)$. To simplify
notations, we sometimes write $\Label{k}(\Cls)$ to denote $\Label{X}(\Cls)$
where $X$ is an arbitrary finite set of size exactly $k$.

\begin{example}
    \label{labelled-well-quasi-ordered:ex}
    The class of all finite paths is \kl{well-quasi-ordered} but not
    \kl{$\infty$-well-quasi-ordered}, while the class of all finite cycles is
    not \kl{well-quasi-ordered}. 
\end{example}

\begin{remark}
    \label{k-well-quasi-ordered:remark}
    Let $\Cls$ be a class of graphs. 
    We can define $l(\Cls)$ as the set of $k \in \Nat_{\geq 1}$ such that
    $\Cls$ is \kl{$k$-well-quasi-ordered}. 
    It is clear that $l(\Cls)$ is a \kl{downwards closed} subset of $(\Nat_{\geq 1}, \leq)$.
    A reformulation of the Pouzet conjecture (\cref{pouzet1:conj}) is that
    $l(\Cls)$ is either $\emptyset$, $\set{1}$, or $\Nat_{\geq 1}$.
\end{remark}

\subsection{Representing Classes of Graphs}

\AP In order to devise algorithms that input a class of graphs and output
whether it is \kl{$\infty$-well-quasi-ordered}, we need to introduce a
syntactic representation for (some) classes of graphs. In this paper we chose
to use the notion of \kl{bounded linear clique-width}, that serves both as a
constraint on the class that ensures that it is ``well-behaved'' (efficient
evaluation of $\MSO$), and provides a concrete representation of such classes
via so-called \emph{$\MSO$-transductions} of words \cite{COJOGR93,COUR94}.
We assume that the reader is familiar with the basics of automata theory,
in particular the notions of monoid morphisms, idempotents in monoids, monadic
second-order ($\MSO$) logic and first-order ($\FO$) logic over finite words
(see e.g. \cite{THOM97}).

\AP
In the rest of the paper, we will use the letter $\Sigma$ to denote a finite
alphabet. A (simple) \intro{$\MSO$-interpretation} from finite words
($\Sigma^*$) to finite graphs is given by an $\MSO$ formula
$\varphi_{\text{edge}}(x,y)$ that defines the edge relation of a graph, a
domain formula $\varphi_{\text{dom}}(x)$ that defines the domain of the output
graph, and a selection formula $\varphi_{\Delta}$ without free variables, used
to restrict the domain of the interpretation.

\begin{definition}
    \label{mso-interpretable:def}

    Let $I \defined (\varphi_{\text{edge}},
    \varphi_{\text{dom}},\varphi_{\Delta})$ be an \kl{$\MSO$-interpretation}
    from finite words to graphs. The image of a word $w \in \Sigma^*$ is the
    graph $G$ with vertices $\setof{1 \leq i \leq |w|}{ w \models
    \varphi_{\text{dom}}(i)}$, and edges $\set{i,j}$ for each pair $(i,j)$ such
    that $i < j$ and $w$ satisfies $\varphi_{\text{edge}}(i,j)$. The
    \kl(interpretation){image} of $I$ is the collection of $I(w)$ where $w$
    ranges over the words in $\Sigma^*$ that satisfy $\varphi_{\Delta}$, we write
    this image $\msoIm(I)$.

\end{definition}

Let us now illustrate in
\cref{clique-bounded-linear-clique-width:ex,paths-bounded-linear-clique-width:ex}
two classes of graphs  that can be obtained as images of
$\MSO$-interpretations. The first one is \kl{$\infty$-well-quasi-ordered},
while the second one is not even \kl{$2$-well-quasi-ordered}.

\begin{example}
    \label{clique-bounded-linear-clique-width:ex}
    The class of all cliques can be obtained as the image of the following
    $\MSO$-interpretation:
        $\varphi_{\text{dom}}(x) \defined \top$, 
        $\varphi_{\text{edge}}(x,y) \defined x \neq y$,
        and $\varphi_{\Delta} \defined \top$.
\end{example}

\begin{example}
    \label{paths-bounded-linear-clique-width:ex}
    The class of all finite paths can be obtained as the image of the following
    $\MSO$-interpretation:
        $\varphi_{\text{dom}}(x) \defined \top$, 
        $\varphi_{\text{edge}}(x,y) \defined (x \leq y) \wedge
        \forall z.  (x \leq z \leq y) \implies z = x \vee z = y$,
        and $\varphi_{\Delta} \defined \top$.
\end{example}

\AP In the introduction, we have hinted at the fact that classes of \intro{bounded linear
clique-width} are precisely those that can be obtained through
\emph{$\MSO$-transductions} of words \cite{COJOGR93,COUR94}. This however,
should be carefully distinguished from the notion of \kl{$\MSO$-interpretation}
we just introduced. Indeed, there are uncountably many classes of graphs that
have \kl{bounded linear clique-width}, and countably many
\kl{$\MSO$-interpretations}. Formally, a class $\Cls$ of graphs has \kl{bounded
linear clique-width} if there exists a finite alphabet $\Sigma$, and an
$\MSO$-interpretation $I$ such that $\Cls \subseteq \msoIm(I)$.

\AP Let us now argue that, to prove the second Pouzet conjecture
(\cref{pouzet2:conj}) for all classes of \kl{bounded linear clique-width}, it
suffices to consider classes of the form $\msoIm(I)$, where $I$ is an
\kl{$\MSO$-interpretation}. This is done by first remarking that the
\kl{hereditary closure} of a class shares similar properties with respect to
\kl{bounded linear clique-width} and \kl{well-quasi-ordering}
(\cref{hereditary-closure:lemma}). And then remarking that for
\kl{$2$-well-quasi-ordered} classes of graphs that have \kl{bounded linear
clique-width}, there exists an $\MSO$-interpretation $I$ that acts as an
\emph{interpolant} between the class and its \kl{hereditary closure}
(\cref{bd-lcw-2-wqo:thm}). We conclude from these two results the desired
\cref{bd-lcw-2-wqo:cor}. Intuitively, one can use an extra label to encode an
\kl{induced subgraph} which explains \cref{hereditary-closure:lemma}. For the
\cref{bd-lcw-2-wqo:thm}, the idea is to use the fact that \kl{hereditary
classes} of graphs that are \kl{$2$-well-quasi-ordered} can be described using
finitely many forbidden induced subgraphs \cite[Proposition 3]{DRT10}, graphs
that can be encoded directly inside the $\MSO$-interpretation.

\begin{lemma}[restate=hereditary-closure:lemma,name=Hereditary Closure Lemma]
    \label{hereditary-closure:lemma}

    Let $k \in \Nat_{\geq 1}$, and $\Cls$ be a class of graphs that is
    \kl{$(k+1)$-well-quasi-ordered}. Then the \kl{hereditary closure} of $\Cls$
    is \kl{$k$-well-quasi-ordered}.

    \proofref{hereditary-closure:lemma}
\end{lemma}

\begin{lemma}[restate=bd-lcw-2-wqo:thm,name=Sandwich Lemma]
    \label{bd-lcw-2-wqo:thm}
    Let $\Cls$ be a \kl{hereditary class} of graphs that has \kl{bounded linear
    clique-width} and is \kl{$2$-well-quasi-ordered}. Then, there exists an
    \kl{$\MSO$-interpretation} $I$ such that
    $
        \Cls \subseteq \msoIm(I)
             \subseteq \dwclosure[\isubleq]{\Cls} 
    $.
    \proofref{bd-lcw-2-wqo:thm}
\end{lemma}

\begin{corollary}[restate={bd-lcw-2-wqo:cor},name=Generalising to Bounded Linear Clique-Width]
    \label{bd-lcw-2-wqo:cor}
    If the second Pouzet conjecture (\cref{pouzet2:conj}) holds for all
    images of \kl{$\MSO$-interpretations}, then it holds for all classes of graphs
    of \kl{bounded linear clique-width}.
    \proofref{bd-lcw-2-wqo:thm}
\end{corollary}

We can strengthen the statement of \cref{bd-lcw-2-wqo:cor} by remarking that
\emph{reasonable} classes of graphs of \kl{bounded linear clique-width} (that
are \kl(class){hereditary} and \kl{$\infty$-well-quasi-ordered}) are all obtained as
images of \kl{$\MSO$-interpretations}. Indeed, the inclusions of
\cref{bd-lcw-2-wqo:thm} become equalities. 
\section{$\MSO$-Interpretations and Monoids}
\label{bounded-lcw:sec}

\AP In this section, we will develop an automata theoretic approach to the
problem of characterizing images of \kl{$\MSO$-interpretations} that are
\kl{$\infty$-well-quasi-ordered} by the \kl{induced subgraph} relation. The key
ingredient is to use the notion of \emph{factorization forests} \cite{SIMO90}
to provide an alternative syntax of such classes of graphs where \emph{words}
(linear trees of unbounded depth) are replaced by \emph{forests} (branching
trees of bounded depth), over which it is easier to define a
\kl{$\infty$-well-quasi-ordering}. In order to simplify our analysis, we will
now assume that in the \kl{$\MSO$-interpretations}, the universe formula
$\varphi_\Delta$ is always $\top$ and that the domain formula
$\varphi_{\text{dom}}(x)$ is too. This is not an actual restriction, as
witnessed by the following proposition.

\begin{lemma}[restate={removing-delta-dom:lemma}, name={Simple Interpretations}]
    \label{removing-delta-dom:lemma}
    Let $I$ be an \kl{$\MSO$-interpretation}. There exists an
    \kl{$\MSO$-interpretation} $I'$ that does not use $\varphi_\Delta$ and
    $\varphi_{\text{dom}}$, and such that for all $k \in \Nat$, $\msoIm(I)$ is
    \kl{$k$-well-quasi-ordered} if and only if $\msoIm(I')$ is.

    Furthermore, $I'$ is effectively computable from $I$.
    \proofref{removing-delta-dom:lemma}
\end{lemma}

\subsection{Monoid-Labelled Classes of Graphs}
\label{factorization-forests:sec}

\AP Let us recall that a monoid $M$ is a set endowed with a binary operation
that is associative, has an identity element $1_M$. Inside a monoid $M$, an
element $e \in M$ is called \intro{idempotent} if $e^2 = e$. A \intro{morphism}
of monoids $\mu \colon M \to N$ is a map that preserves the multiplication and
the identity element, i.e., $\mu(1_M) = 1_N$ and $\mu(ab) = \mu(a)\mu(b)$ for
all $a,b \in M$. 

\AP It follows from classical results in automata theory that one can associate
to every \kl{$\MSO$-interpretation} $I$ a finite \kl{monoid} $M$, a morphism
$\mu \colon \Sigma^* \to M$ such that evaluating the $\MSO$-interpretation on a
word $w \in \Sigma^*$ can be rephrased in terms of selecting images of
decomposition of $w$ though $\mu$ \cite{EILE74}. Rather than
working directly with these monoids, we propose the following notion of
\kl{monoid-labelled graphs} that hides the technicalities of the word
construction, and provides a bounded-depth tree-like description of graphs of
\kl{bounded linear clique-width}. The main idea of this new combinatorial
object is to recover \emph{compositional behaviors} from an
\kl{$\MSO$-interpretation}.

\begin{definition}
    \label{monoid-labelled-graph:def}

    Let $M$ be a finite monoid. A \intro{monoid-labelled graph} $\aMgraph$ over $M$ is a
    tuple $\aMgraph = (m, V, E, \vlbl)$ where $m \in M$, $V$ is a set of vertices, $E
    \colon V^2 \to (M^2 \to \Bool)$, and $\vlbl \colon V \to M^2$,
    such that
    for all $x \in V$, $(l_x, r_x) = \vlbl(x)$, we have $l_x r_x = m$.

    We say that such a graph \intro(mlg){evaluates} to $m$.
\end{definition}

\begin{figure}
    \centering
    \begin{tikzpicture}
        \begin{scope}
            \draw (0,0) rectangle (4,1);
            \node[label=above:$m$] at (2,1) {};
            \node[label=below:$\aMgraph$] at (2,0) {};
            \node[yshift=0.25cm,leftlabel] at (1,0.5) {$l_x$};
            \node[yshift=0.25cm,rightlabel] at (3,0.5) {$r_x$};
            \draw (0,1.1) -- (4,1.1);
            \draw[leftlabel] (0,0.5) -- (2,0.5);
            \draw[rightlabel] (2,0.5) -- (4,0.5);
            \node[fill,circle,minimum size=0.1cm,inner sep=0pt,label=below:$x$] at (2,0.5) {};
        \end{scope}
        \begin{scope}[xshift=6cm]
            \draw (0,0) rectangle (4,1);
            \node[label=above:$amb$] at (2,1) {};
            \node[label=below:$a\aMgraph{}b$] at (2,0) {};
            \node[yshift=0.25cm,leftlabel] at (1,0.5) {$al_x$};
            \node[yshift=0.25cm,rightlabel] at (3,0.5) {$r_xb$};
            \draw (0,1.1) -- (4,1.1);
            \draw[leftlabel] (0,0.5) -- (2,0.5);
            \draw[rightlabel] (2,0.5) -- (4,0.5);
            \node[fill,circle,minimum size=0.1cm,inner sep=0pt,label=below:$x$] at (2,0.5) {};
        \end{scope}
    \end{tikzpicture}
    \caption{An abstract representation of a \kl{monoid-labelled graph} $\aMgraph$,
        that \kl(mlg){evaluates} to the monoid element $m$,
        and one of its vertices $x$.
        Letting $(l_x, r_x) \in M^2$ be the label of $x$ in $\aMgraph$,
        we provide a graphical representation of the equation $l_x m_x r_x = m$.
        We also illustrate how the monoid $M$ can be used to act on the left and on the right
        on the graph $\aMgraph$ by drawing the same abstract representation of
        $a \aMgraph b$.
    }
    \label{monoid-labelled-graph:fig}
\end{figure}
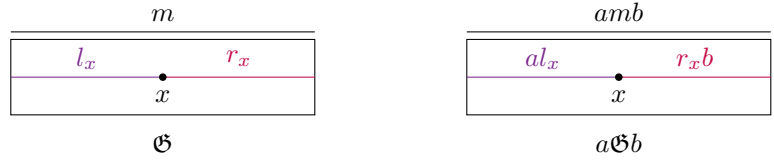

\AP We provide in \cref{monoid-labelled-graph:fig} a useful graphical
representation of a \kl{monoid-labelled graph}, that abstracts the edge
relation and focuses on the labels of the whole graph and its vertices. Using
this representation, it should be easier to visualize the graph operations that
will be defined, and for instance how one can multiply on the left and on the
right a \kl{monoid-labelled graph} $\aMgraph$ by elements $a,b \in M$: the
graph has the same vertices as $\aMgraph$, the graph \kl(mlg){evaluates} to $a
m b$ where $m$ \kl(mlg){evaluates} to $m$, the label of a vertex $x$ is updated
as $a \vlbl(x) b \defined (a\vlbl(x)_1, \vlbl(x)_2 b)$, and given vertices
$x,y$, the edge relation is updated as follows $aEb (s,t) \defined E(as, tb)$.

\AP Given a \kl{monoid-labelled graph} $\aMgraph$, one can construct is
\intro{monoid downcasting} as a \kl{labelled graph} $\intro*\mcast{\aMgraph}$
by keeping the labelling of the graph intact, and defining the following edge
structure: there is an edge between $x$ and $y$, if and only if $E(x,y)(\unit,
\unit) = \top$. When it is clear from the context, we can implicitly forget
about the labels of the \reintro{monoid downcasting}, and consider it directly
as an unlabelled graph.

\begin{figure}
    \centering
    \begin{tikzpicture}
        \begin{scope}
        \draw (0,0) rectangle (4,1);
        \node[label=above:$m_1$] at (2,1) {};
        \node[label=below:$\aMgraph_1$] at (2,0) {};
        \node[yshift=0.25cm,A2] at (1,0.5) {$l_x$};
        \node[yshift=0.25cm,A3] at (3,0.5) {$r_x$};
        \draw (0,1.1) -- (4,1.1);
        \draw[A2] (0,0.5) -- (2,0.5);
        \draw[A3] (2,0.5) -- (4,0.5);
        \node[A4,fill,circle,minimum size=0.1cm,inner sep=0pt,label=below:$x$] at (2,0.5) {};
        \end{scope}
        \begin{scope}[xshift=4cm]
        \draw (0,0) rectangle (4,1);
        \node[label=above:$m_2$] at (2,1) {};
        \node[label=below:$\aMgraph_2$] at (2,0) {};
        \node[yshift=0.25cm,A2] at (1,0.5) {$l_y$};
        \node[yshift=0.25cm,A3] at (3,0.5) {$r_y$};
        \draw (0,1.1) -- (4,1.1);
        \draw[A2] (0,0.5) -- (2,0.5);
        \draw[A3] (2,0.5) -- (4,0.5);
        \node[A4,fill,circle,minimum size=0.1cm,inner sep=0pt,label=below:$y$] at (2,0.5) {};
        \end{scope}
        \begin{scope}[xshift=2cm,yshift=-2.2cm]
        \draw (0,0) rectangle (4,1);
        \node[label=above:$m_1 m_2$] at (2,1) {};
        \node[label=below:$\aMgraph_1 \bp \aMgraph_2$] at (2,0) {};
        \node[yshift=0.25cm,A2] at (1,0.5) {$m_1l_y$};
        \node[yshift=0.25cm,A3] at (3,0.5) {$r_y$};
        \draw (0,1.1) -- (4,1.1);
        \draw[A2] (0,0.5) -- (2,0.5);
        \draw[A3] (2,0.5) -- (4,0.5);
        \node[A4,fill,circle,minimum size=0.1cm,inner sep=0pt,label=below:$y$] at (2,0.5) {};
        \end{scope}
    \end{tikzpicture}
    \caption{An abstract representation of the
        \kl{binary product}
        of two \kl{monoid-labelled graphs} $\aMgraph_1$ and $\aMgraph_2$ via
        an \kl{edge selector} $P_{\text{edge}} \subseteq M^3$.
        We have selected two vertices $x$ and $y$ in the two graphs, and illustrated
        how the label of $y$ in $\aMgraph_1 \bp \aMgraph_2$ is computed.
        We graphically observe see how the edge relation is updated in the product:
        the edge function between $x$ and $y$ in $\aMgraph_1 \bp \aMgraph_2$
        in a context $a,b \in M^2$
        is defined as $E(x,y)(a,b) \defined (a l_x, r_x l_y, r_y b) \in P_{\text{edge}}$.
    }
    \label{binary-product:fig}
\end{figure}
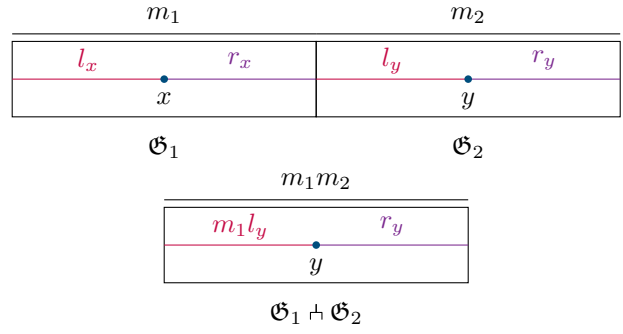

\AP Let us now propose two composition operations on \kl{monoid-labelled
graphs}. These operations will need an additional input as a subset
$P_{\text{edge}} \subseteq M^3$ which we call an \intro{edge selector}.
However, in the combinatorial analysis, this subset will be \emph{fixed}, and we
will therefore omit it from the notation to avoid cluttering.
There is a graphical representation of \cref{binary-product:def} in
\cref{binary-product:fig}.

\begin{definition}
    \label{binary-product:def}
    Let $M$ be a finite monoid, 
    $P_{\text{edge}} \subseteq M^3$ be an \kl{edge selector},
    and let $\aMgraph_1$ and $\aMgraph_2$ be two
    \kl{monoid-labelled graphs}. The \intro{binary product} of $\aMgraph_1$ and
    $\aMgraph_2$ with respect to $P_{\text{edge}}$
    written $\aMgraph_1 \bp \aMgraph_2$ is the \kl{monoid-labelled graph}
    obtained as follows:
    \begin{itemize}
        \item Its vertices are the disjoint union of the vertices of $\aMgraph_1$
            and $\aMgraph_2$,
        \item It \kl(mlg){evaluates} to the product $m_1 m_2$
            of the respective \kl(mlg){evaluations}
            of $\aMgraph_1$ and $\aMgraph_2$,
        \item The label of a vertex $x \in V_1$ is $(l_x, r_x m_2)$
            where $(l_x, r_x) = \vlbl_1(x)$.
        \item The label of a vertex $y \in V_2$ is $(m_1 l_y, r_y)$
            where $(l_y, r_y) = \vlbl_2(y)$.
        \item The edge relation between elements in $V_1$
            is obtained as the edge relation of $\aMgraph_1$ multiplied on the right by $m_2$,
            similarly for the edge relation between elements in $V_2$ 
            (multiplied on the left by $m_1$).
            An edge between $x \in V_1$ and $y \in V_2$ exists in a context $a,b \in M^2$
            if and only if $(a l_x, r_x l_y, r_y b) \in P_{\text{edge}}$.
    \end{itemize}
\end{definition}

\AP A second important operation, which is the key ingredient in providing
\emph{bounded depth} expressions for some classes of graphs, is the
\kl{idempotent product} of a \emph{sequence} of \kl{monoid-labelled graphs}.
Remark that the \kl{binary product} is associative, and that therefore the
\emph{product} of a sequence of \kl{monoid-labelled graphs} is well-defined. 

\begin{definition}
    \label{idempotent-product:def}
    Let $M$ be a finite monoid,
    $P_{\text{edge}} \subseteq M^3$ be an \kl{edge selector},
    and let $\seqof{\aMgraph_i}{1 \leq i \leq n}$ be a family of
    \kl{monoid-labelled graphs} all \kl(mlg){evaluating} to the same \kl{idempotent} monoid element. The \intro{idempotent product} of the family 
    is written $\ep \seqof{\aMgraph_i}{1 \leq i \leq n}$ and is the \kl{monoid-labelled graph}
    and is simply the product of the family.
\end{definition}

\AP Let us now introduce the class of \kl{regular monoid-labelled graphs} that
can be obtained by iteratively applying these operations, which will be the
combinatorial counterpart of graphs of \kl{bounded linear clique-width}. In
particular, \kl{regular monoid-labelled graphs} will have \emph{syntax trees}
of \emph{bounded depth}.

\begin{definition}
    \label{regular-monoid-labelled-graphs:def}
    Let $M$ be a finite monoid, and $P_{\text{edge}} \subseteq M^3$ be an \kl{edge selector}.
    The class of \intro{regular monoid-labelled} graphs is inductively defined as follows
    \begin{itemize}
        \item $\RegMG{0}$ is the class of \kl{monoid-labelled graphs} with a single vertex,
        \item $\RegMG{k+1}$
            contains all the graphs $\aMgraph_1 \bp \aMgraph_2$ 
            where $\aMgraph_1, \aMgraph_2 \in \RegMG{k}$,
            and
             $\ep \seqof{\aMgraph_i}{1 \leq i \leq n}$
            where $\seqof{\aMgraph_i}{1 \leq i \leq n}$
            is a family of \kl{monoid-labelled graphs} 
            in $\RegMG{k}$
            that all \kl(mlg){evaluate} to the same \kl{idempotent} element.
    \end{itemize}
\end{definition}

The following theorem is a direct consequence of the classical correspondence
between monadic second order logic and monoids in automata theory, combined
with the factorization forest theorem of \cite{SIMO90}, proving that words in a
given monoid $M$ can be factorized into a forest of bounded depth using binary
products and idempotent products. 

\begin{theorem}[{\cite{SIMO90}}]
    \label{factorization-forest-theorem:thm}
    Let $\Cls$ be a \kl{hereditary class} of graphs.
    The following are equivalent:
    \begin{itemize}
        \item $\Cls$ is the image of some \kl{$\MSO$-interpretation} (without $\varphi_{\Delta}$ and $\varphi_{\text{dom}}$),
        \item There exists a finite monoid $M$,
            a depth $d \in \Nat$, and an \kl{edge selector} $P_{\text{edge}} \subseteq M^3$
            such that
            $\Cls$ is
            the \kl{monoid downcasting} of the class of \kl{regular monoid-labelled graphs}
            $\RegMG{d}$, ignoring the labels on the vertices.
    \end{itemize}
    Furthermore, explicit conversions between the two representations are effective.
\end{theorem}

\subsection{Ordering Regular Monoid-Labelled Graphs}
\label{ordering-regular-monoid-labelled-graphs:sec}

\AP In this section, we fix a finite monoid $M$, an \kl{edge selector}
$P_{\text{edge}} \subseteq M^3$, and our goal is to understand what kind of
orderings can be placed on the class of \kl{regular monoid-labelled graphs}
$\RegMG{d}$, for a fixed $d$, so that it is \kl{$\infty$-well-quasi-ordered}.
Such an ordering can be introduced by induction on $d$. Since $\RegMG{0}$ is a
finite set, it is always \kl{$\infty$-well-quasi-ordered}. For the inductive
step, because finite products of \kl{well-quasi-orderings} remain
\kl{well-quasi-ordered}, the only difficulty lies in understanding how to
provide an ordering on \kl{idempotent products} of graphs.

\begin{figure}
    \centering
    \begin{tikzpicture}[
            firstnode/.style={fill,circle,minimum size=0.1cm,inner sep=0pt},
            secondnode/.style={circle,draw,minimum size=0.1cm,inner sep=0pt},
        ]
        \begin{scope}
            \draw (0,0) rectangle (5,1);
            \foreach \i in {1,2,3,4}
            {
                \draw (\i,0) -- (\i,1);
            }
            \foreach \i in {1,2,3,4,5}
            {
                \node[label=below:$\aMgraph$] at (\i-0.5,0) {};
                \node[firstnode] (F\i) at (\i-0.5,0.7) {};
                \node[secondnode] (S\i) at (\i-0.5,0.3) {};
                \draw (\i-1,1.1) -- (\i,1.1);
                \node[label=above:$e$] at (\i-0.5,1) {};
            }
            \foreach \i in {1,2,3,4}
            {
                \draw (F\i) -- (S\the\numexpr\i+1\relax);
            }
        \end{scope}
        \begin{scope}[xshift=6cm]
            \draw (0,0) rectangle (5,1);
            \foreach \i in {1,2,3,4}
            {
                \draw (\i,0) -- (\i,1);
            }
            \foreach \i in {1,2,3,4,5}
            {
                \node[label=below:$\aMgraph$] at (\i-0.5,0) {};
                \node[firstnode] (F\i) at (\i-0.5,0.7) {};
                \node[secondnode] (S\i) at (\i-0.5,0.3) {};
                \draw (\i-1,1.1) -- (\i,1.1);
                \node[label=above:$e$] at (\i-0.5,1) {};
            }
            \foreach \i in {1,2,3,4}
            {
                \draw (F\i) -- (F\the\numexpr\i+1\relax);
            }
        \end{scope}
    \end{tikzpicture}
    \caption{The importance of the \kl{edge selector}
        can be viewed on the following \kl{idempotent product}
        of graphs. On the left, the edge selection
        is such that the \kl{idempotent product} of the graph $\aMgraph$
        with itself leads to a collection of disjoint paths,
        while on the right, the edge selection is such that the same product
    leads to a long path with extra isolated vertices.
    }
    \label{idempotent-product-paths-disj:fig}
\end{figure}
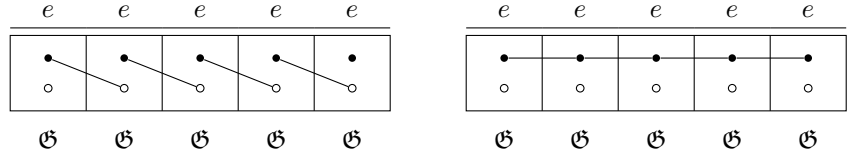

\AP We draw in \cref{idempotent-product-paths-disj:fig} two examples of
\kl{idempotent products} of a repetition of a single \kl{monoid-labelled
graphs} that \kl(mlg){evaluates} to the \kl{idempotent} monoid element $e$. In
this diagram, we draw the edges in the \kl{monoid downcasting} of the resulting
graph. Because the element $e$ is an \kl{idempotent}, the presence of an edge
between two vertices $x$ and $y$ that are not in consecutive parts of the
product only depends on the labels of $x$ and $y$, here we decided that this
will always be a non-edge. On consecutive products, it can be that some edges
are created, based on the specific choice of an \kl{edge selector}
$P_{\text{edge}}$. On the left example, the class of graphs obtained by
\kl{monoid downcasting} of the \kl{idempotent product} is a collection of
disjoint paths, which is a \kl{$\infty$-well-quasi-ordered} class of graphs for
the \kl{induced subgraph} relation. However, on the right example, the class of
graphs obtained by \kl{monoid downcasting} of the \kl{idempotent product}
contains arbitrarily long induced paths, and is therefore not
\kl{$\infty$-well-quasi-ordered}.

\AP This motivates our definition of \kl{good forest paths} and \kl{bad forest
paths}, which is the key distinguishing criterion between classes that will be
\kl{$\infty$-well-quasi-ordered}.

\begin{definition}
    \label{good-forest-paths:def}
    Let $\seqof{\aMgraph_i}{1 \leq i \leq n}$ be a sequence of \kl{monoid-labelled graphs}
    that all evaluate to some \kl{idempotent} element $e$.
    The sequence is a \intro{good forest path} if for all context $a,b \in M^2$,
    there exists another sequence
    $\seqof{\aMgraph[H]_j}{1 \leq j \leq k}$ evaluating to $e$, 
    and an \kl(graph){embedding} $h \colon G \to H$
    where $G = \mcast{a \ep \seqof{\aMgraph_i}{1 \leq i \leq n} b}$
    and $H = \mcast{a \ep \seqof{\aMgraph[H]_j}{1 \leq j \leq k} b}$,
    such that
    \begin{itemize}
        \item $h$ preserves the labels of the vertices in the graphs $\aMgraph_i$ and $\aMgraph[H]_j$,
        \item the image of $\aMgraph_1$ is contained in $\aMgraph[H]_1$,
        and the image of $\aMgraph_n$ is contained in $\aMgraph[H]_k$,
        \item there exists $1 \leq j \leq k$ such that the image of $h$ does not intersect $\aMgraph[H]_j$.
    \end{itemize}
    We say that the \kl{good forest path} is \intro{split}
    by the sequence $\seqof{\aMgraph[H]_j}{1 \leq j \leq k}$ and 
    \kl(graph){embedding} $h$.
    A \intro{bad forest path} is a forest path that is not a \kl{good forest path}.
\end{definition}

\AP Typically, \kl{good forest paths} can be \kl{split} into smaller
independent parts. Note that the \kl{split} can reorder the nodes in some
arbitrary fashion, not even respecting the order of the nodes in the original
forest. Our first result is that, unsurprisingly, the existence of long \kl{bad
forest paths} provides counter examples to being
\kl{$\infty$-well-quasi-ordering}. This result crucially relies on the ability
to place \emph{labels} on the vertices of a graph to encode the original
\kl{monoid-labelled graph} it was obtained from. This explains why the actual
lemma bounds precisely the number of colors used.

\begin{lemma}[restate={bad-forests-paths-imply-antichains:lemma}, name={Antichains of Bad Forest Paths}]
    \label{bad-forests-paths-imply-antichains:lemma}
    Let $d \in \Nat$. Assume that \kl{bad forest paths} of arbitrarily large length
    in $\RegMG{d}$.
    Then, the class of \kl{monoid downcasting} of the graphs in $\RegMG{d}$
    is not \kl{$k$-well-quasi-ordered} for the \kl{induced subgraph} relation,
    where $k = |M|^2 \times 3$.
    \proofref{bad-forests-paths-imply-antichains:lemma}
\end{lemma}

\AP For the converse implication, we will use the following remark on \kl{good
forest paths}, they can be \kl{split} without requiring new exotic
\kl{monoid-labelled graphs} or complex \kl(graph){embeddings}: it suffices to
repeat the original sequence three times to obtain a decomposition.

\begin{lemma}[restate={good-forest-paths-decomp:lemma}, name={Canonical Good Decomposition}]
    \label{good-forest-paths-decomp:lemma}
    Let $\seqof{\aMgraph_i}{1 \leq i \leq n}$ be a \kl{good forest path}.
    Then,
    for all $a,b \in M^2$, there exists an \kl(graph){embedding}
    $h \colon G \to H$
    where
    $G = \mcast{a \ep \seqof{\aMgraph_i}{1 \leq i \leq n} b}$
    $H = \mcast{a (\ep \seqof{\aMgraph_i}{1 \leq i \leq n})^3 b}$
    satisfying the requirements of 
    \cref{good-forest-paths:def}.
    \proofref{good-forest-paths-decomp:lemma}
\end{lemma}

Using the decomposition provided in \cref{good-forest-paths-decomp:lemma}, we
can now prove by induction on $d$ that the \kl{monoid downcastings} of the
graphs in $\RegMG{d}$ form a \kl{$\infty$-well-quasi-ordered} class of graphs
for the \kl{induced subgraph} relation. The only non-trivial case being the
\kl{idempotent products}, which is handled by \kl{splitting} the \kl{good
forest paths} and then using the usual notion of (well-quasi-)ordering on
sequences (the \kl{subword ordering}) due to Higman \cite{HIG52}.

\begin{lemma}[restate={bounded-forest-paths:theorem}, name={Characterising Well-Quasi-Orders}]
    \label{bounded-forest-paths:theorem}
    Let $d \in \Nat$. The following properties are equivalent:
    \begin{enumerate}
        \item There is a bound on the length of \kl{bad forest paths}
            in $\RegMG{k}$ for all $k \leq d$.
        \item The class of \kl{monoid downcastings} of $\RegMG{d}$
            is \kl{labelled-well-quasi-ordered} for the \kl{induced subgraph} relation.
    \end{enumerate}
    \proofref{bounded-forest-paths:theorem}
\end{lemma}

Let us underline the fact that no cleverness is needed in the proof of
\cref{bounded-forest-paths:theorem}: it simply builds a
\kl{well-quasi-ordering} inductively by applying product and subword
constructions adapted to the \kl{monoid-labelled graphs}.
To turn \cref{bounded-forest-paths:theorem} into a final theorem, one lacks a
decision procedure, which we provide in the following lemma. The main idea is
that the collection of \kl{bad forest paths} is definable in $\MSO$, and
therefore the existence of a bound amounts to checking whether a regular
language is bounded, which is decidable.

\begin{lemma}[restate={decide-bounded-forest-paths:lemma}, name={Deciding Bounded Forest Paths}]
    \label{decide-bounded-forest-paths:lemma}
    Let $d \in \Nat$. It is decidable whether
    $\RegMG{d}$ contains \kl{bad forest paths} of unbounded length.
    \proofref{decide-bounded-forest-paths:lemma}
\end{lemma}

As a consequence, we can now obtain our main theorem, characterizing images of
\kl{$\MSO$-interpretations} that are \kl{$\infty$-well-quasi-ordered}. Because
of \cref{removing-delta-dom:lemma} and \cref{bd-lcw-2-wqo:cor}, this
characterization can be lifted to answer
\cref{pouzet2:conj} positively in the case of graphs with \kl{bounded linear
clique-width}, dropping the effective computability of the bound $k$. Remark
that if the bound $k$ happened to be $2$ in \cref{main-theorem:thm}, this would
imply \cref{pouzet1:conj} for classes of graphs of \kl{bounded linear clique-width}.

\begin{theorem}[restate={main-theorem:thm}, name={Main Result}]
    \label{main-theorem:thm}
    Let $\Cls$ be the image of an \kl{$\MSO$-interpretation} $I$.
    There exists a computable $k \in \Nat_{\geq 1}$ such that the following are equivalent:
    \begin{enumerate}
        \item \label{main-theorem-kwqo:item}
            $\Cls$ is \kl{$k$-well-quasi-ordered} by the \kl{induced subgraph} relation,
        \item \label{main-theorem-lwqo:item}
            $\Cls$ is \kl{$\infty$-well-quasi-ordered} by the \kl{induced subgraph} relation,
        \item \label{main-theorem-wwqo:item}
            $\Cls$ is \kl{labelled-well-quasi-ordered} by the \kl{induced subgraph} relation.
    \end{enumerate}
    Furthermore, these properties are decidable.
\end{theorem}
\begin{proof}
    First, remark that implications \cref{main-theorem-wwqo:item}
    $\Rightarrow$ \cref{main-theorem-lwqo:item} $\Rightarrow$
    \cref{main-theorem-kwqo:item} are true for all $k \in \Nat_{\geq 1}$. 

    Let $\Cls$ be
    the image of an \kl{$\MSO$-interpretation} $I$. 
    Because of \cref{removing-delta-dom:lemma}, we can assume that
    the interpretation does not use the
    domain formula $\varphi_{\text{dom}}$ nor the
    selection formula $\varphi_{\Delta}$.
    Let $M$ be the monoid
    associated to $I$ using \cref{factorization-forest-theorem:thm}, and $d$ be
    the corresponding depth of expressions. Let $k \defined 3 \times |M|$.
    Assume that $\Cls$ is \kl{$k$-well-quasi-ordered}. Because of
    \cref{bad-forests-paths-imply-antichains:lemma}, this implies that there is
    a bound on the \kl{bad forest paths} in $\RegMG{d}$. Thanks to
    \cref{bounded-forest-paths:theorem}, this implies that $\Cls$ is
    \kl{labelled-well-quasi-ordered}.
\end{proof}

\begin{corollary}[restate={main-corollary:cor}, name={Second Pouzet Conjecture}]
    \label{main-corollary:cor}
    Let $\Cls$ be a class of graphs having \kl{bounded linear clique-width}.
    Then, $\Cls$ is \kl{$\infty$-well-quasi-ordered} by the \kl{induced subgraph} relation
    if and only if it is
    \kl{labelled-well-quasi-ordered} by the \kl{induced subgraph} relation.
\end{corollary}
\section{Totally Ordered Monoids}
\label{totally-ordered-monoids:sec}

\AP This section is devoted to comparing \cref{main-theorem:thm} with
previously existing results. In particular, we will try to understand the work
of Daligault, Rao, and Thomass\'e in the language of finite monoids, where we
will find connections that may be of interest to the community. Let us first
re-introduce the notion of \kl{tree-model} that was used as a witness that a
class of graphs has \kl{bounded clique-width} in \cite{DRT10}. Instead of using
the notion of \emph{relabel function}, which are endofunctions from a finite
set $Q$ closed under composition, we directly work on an abstract finite monoid
$M$.

\begin{definition}[{\cite{DRT10}}]
    Let $M$ be a finite monoid.
    A \intro{tree model} is a tuple $\aTreeModel = (T, \leq, \mu, \lambda)$ where
    $T$ is a tree represented with the ancestor relation $\leq$,
    $\mu \colon V(T) \to M^2$ is a selection function,
    and $\lambda \colon E(T) \to M$ is a morphism that maps every
    edge (for the parent relation) to an element of a monoid $M$.
\end{definition}

\AP A \kl{tree-model} can be mapped to a graph $G$ by considering the leaves of
the tree as vertices, and adding an edge between two leaves $x$ and $y$ as
follows: compute $l$ their least common ancestor, $m_x = \lambda(x,l)$ and $m_y
= \lambda(y,l)$, and check whether $(m_x, m_y) \in \mu(l)$. Let us note this
function $\tmeval$. It is a classical result that a class of graphs $\Cls$ has
\kl{bounded clique-width} if and only if there exists a monoid $M$ such that
$\Cls$ is contained in the image of $\tmeval$ \cite{DRT10}.

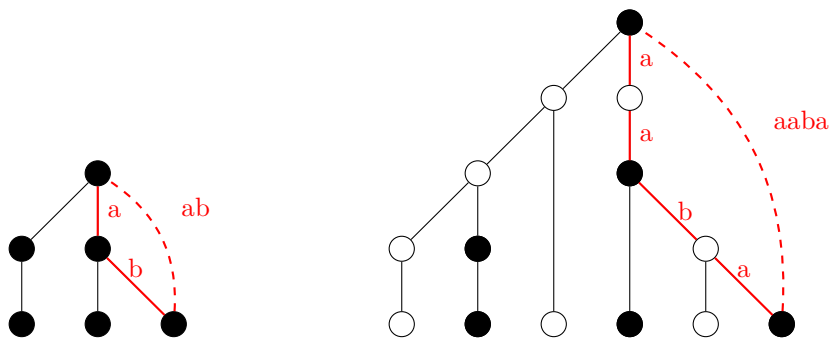
\begin{figure}
    \centering
    \begin{tikzpicture}
        \foreach[count=\i] \x/\y in {0/0,0/1,0/2,1/0,1/1,2/1} {
            \node[draw, circle,fill=black] (t\i) at (\y,\x) {};
        }
        \foreach \i/\j in {4/1,5/2,5/3,6/4,6/5} {
            \draw (t\j) -- (t\i);
        }
        \begin{scope}[xshift=5cm]
        \foreach[count=\i] \x/\y in {0/0,0/1,0/2,0/3,0/4,0/5,1/0,1/1,1/4,2/1,2/3,3/2,3/3,4/3} {
            \node[draw, circle] (h\i) at (\y,\x) {};
        }
        \foreach \i/\j in {7/1,8/2,9/5,9/6,10/7,10/8,11/4,11/9,12/10,12/3,13/11,14/12,14/13} {
            \draw (h\j) -- (h\i);
        }
        \end{scope}
        \foreach \j in {2,4,6,8,11,14} {
            \node[draw, circle, fill=black] at (h\j) {};
        }
        \draw[red,thick] (t3) -- node[midway, above] {b} 
                         (t5) -- node[midway, right] {a}
                         (t6);
        \draw[red,thick,dashed] (t3) to[bend right] node[midway, above right=0.2cm] {ab} (t6);

        \draw[red,thick] (h6)  -- node[midway, above] {a} 
                         (h9)  -- node[midway, right] {b}
                         (h11) -- node[midway, right] {a}
                         (h13) -- node[midway, right] {a}
                         (h14);
        \draw[red,thick,dashed] (h6) to[bend right] node[midway, above right=0.2cm] {aaba} (h14);

    \end{tikzpicture}
    \caption{Embeddings of tree models respecting the monoid structure. We depict
        on the left a tree, where some edges are labelled with elements of a monoid $M$.
        On the right, we provide a tree in which the left tree can embed,
        respecting the least common ancestor relation, images of the left tree
        are represented as black nodes. Because the embedding respects the monoid
        structure, we conclude that $ab = aaba$
        since the image of the red path on the left is the red path on the right,
        and the product of the labels on the path is preserved by the 
        \kl{tree-model embedding relation}.
    }
\end{figure}

\AP The key idea of \cite{DRT10} was to study a variation of the \intro{Kruskal
embedding} of trees \cite{KRU72}, suitably adapted to be compatible with the
edge labelling $\lambda$. Such a \intro{tree-model embedding} is defined as a
map $h$ from vertices to vertices, respecting the ancestor relation, the colors
on the vertices, and such that for all edge $x \leq y$ in the input
\kl{tree-model}, the label $\lambda(x,y) = \lambda(h(x), h(y))$, the latter
being well-defined since $h(y)$ is an ancestor of $h(x)$. Alternatively, one
can think of replacing an edge in the tree by a path such that the products of
the labels along the path is the same as the label of the original edge. We
write $T \intro*\tmleq  T'$ if there exists a \kl{tree-model embedding} from
$T$ to $T'$.

\AP The interest of such a \kl{tree-model embedding} is that the function
$\tmeval$ is now an \intro{order reflection} from the class of \kl{tree-models}
equipped with the \kl{tree-model embedding} relation to its image equipped with
the \kl{induced subgraph} relation. Let us recall that an \reintro{order
reflection} from $(A, \leq)$ to $(B, \sqsubseteq)$ is a map $f \colon A \to B$
such that $f(x) \sqsubseteq f(y)$ implies $x \leq y$ for all $x,y \in A$. If
$f$ is an \reintro{order reflection} from $(A, \leq)$ to $(B, \sqsubseteq)$,
and $(B, \sqsubseteq)$ is a \kl{well-quasi-ordering}, then $(A, \leq)$ is one
too.

\AP For some choice of monoid $M$, the \kl{tree-model embedding}
relation is \emph{not} a \kl{$\infty$-well-quasi-ordering}. The main result of
Daligault, Rao and Thomass\'e reads as follows: one can decide, given $M$,
whether the class of graphs generated by the \kl{tree-models} is
\kl{$\infty$-well-quasi-ordered}, and the latter is equivalent to being
\kl{$2$-well-quasi-ordered} \cite[Theorem 3]{DRT10}.

\AP Let us first remark that \cite[Theorem 3]{DRT10} cannot be used to decide
if a class $\Cls$ of graphs with \kl{bounded clique-width} (or even,
\kl{bounded linear clique-width}) is \kl{$\infty$-well-quasi-ordered}. The
reason is that the class $\Cls$ may be a \emph{subset} of the class of graphs
generated by the \kl{tree-models}, and therefore may be
\kl{$\infty$-well-quasi-ordered} even when the bigger class is not.\footnote{It is not
    clear whether there is an equivalent to our
    \cref{bd-lcw-2-wqo:cor} in the context of \cite{DRT10}.}
Furthermore, the proof of the theorem relies on a \emph{minimal bad sequence
argument} \cite{NASH65}, and non-trivial combinatorial arguments on the
structure of relabeling functions.

\subsection{Gap Embedding and Well-Quasi-Ordering of Graphs}
\label{gap-embedding:sec}

\AP In their seminal paper, \cite{DRT10} introduced the notion of ordering of
relabeling functions defined as follows: given $f,g \colon Q \to Q$ be two
endofunctions on a finite set $Q$, they define $f \dalileq g$ as $\image{f}
\subseteq \image{f \circ g}$, and study submonoids of endofunctions on $Q$ that
are \emph{totally ordered} for this relation. However, it seems that this
notion of ordering is an artifact of a more general phenomenon that can be
understood using classical tools from monoid theory called the \emph{Green
Relations} \cite{COLC11}. Let $M$ be a finite monoid and $m \in M$ be an
element. We denote by $\intro*\jideal{m}$ the \intro{bilateral ideal} of $m$,
i.e., the set of elements $xmy$ where $x,y$ ranges in $M$. Similarly, we define
the \intro{left ideal} $\intro*\lideal{m}$ and the \intro{right ideal}
$\intro*\rideal{m}$ respectively as the sets of elements $xm$ and $my$ where
$x,y$ ranges in $M$.

The following lemma provides a rephrasing of the conditions of \cite{DRT10} in
in terms of monoid ideals and therefore generalizes this characterization to
any finite monoid.

\begin{lemma}[restate={equivalent-total-orderings:lem},name={Green Formulations of Total Orderings}]
    \label{equivalent-total-orderings:lem}
    Let $M$ be a finite monoid.
    The following are equivalent:
    \begin{enumerate}
        \item \label{equivalent-total-orderings-rideal:item}
            $M$ is totally ordered by the relation defined by
            $x \leq y$ if and only if 
            $\rideal{x} \subseteq \rideal{x y}$,
        \item \label{equivalent-total-orderings-lideal:item}
            $M$ is totally ordered by the relation defined by
            $x \leq y$ if and only if 
            $\lideal{x} \subseteq \lideal{y x}$,
        \item \label{equivalent-total-orderings-jideal:item}
            $M$ is totally ordered by the relation defined by
            $x \leq y$ if and only if 
            $\jideal{x} = \jideal{x y} = \jideal{y x}$,
        \item \label{equivalent-total-orderings-jideal2:item}
            The \kl{bilateral ideals} of $M$ are totally ordered
            for inclusion, and for all $x,y \in M$,
            $\jideal{x y} = \jideal{x} \cap \jideal{y}$.
    \end{enumerate}
    Whenever one of these conditions is satisfied, all the defined preorders
    coincide.
    Furthermore, if $M$ is a submonoid of the endofunctions of a finite set $Q$,
    then the condition is equivalent to the condition that $M$ is totally ordered
    for the relation $\dalileq$.

    \proofref{equivalent-total-orderings:lem}
\end{lemma}

Let us now introduce a definition for so-called ``totally ordered monoids''
that is based on \cref{equivalent-total-orderings-jideal:item}. We provide two
examples of such monoids: groups in \cref{totally-ordered-groups:ex} and band
monoids \cref{totally-ordered-band:ex}. These two examples are here to
illustrate that the condition of being a \kl{totally ordered monoid} is not too
restrictive.

\begin{definition}
    A monoid $M$ is \intro{totally ordered}
    if for all $a,b \in M$,
    either
    $\jideal{ab} = \jideal{a}$ or $\jideal{ab} = \jideal{b}$.
    We write $a \intro*\jleq b$ if $\jideal{a} \subseteq \jideal{b}$,
    and $a \intro*\jequiv b$ if $\jideal{a} = \jideal{b}$.
\end{definition}

\begin{example}[restate={totally-ordered-groups:ex}]
    \label{totally-ordered-groups:ex}
    Every group $(G, \cdot)$ is a \kl{totally ordered monoid}.
\end{example}

\begin{example}[restate={totally-ordered-band:ex}]
    \label{totally-ordered-band:ex}
    Let $M$ be a monoid, i.e., a monoid where for all $a,b \in M$,
    there exists $c \in M$ such that $a = c b c$. Then $M$ is a \kl{totally
    ordered monoid}.
\end{example}

Totally ordered monoids enjoy the following \emph{cancellation property}, that
was noticed by \cite{DRT10}, and powers their combinatorial results. In the
upcoming \cref{gap-embedding-tree-model:thm} we only use the cancellation
property on the left (as in \cite{DRT10}), but cancellation on the right will be
crucial in the proof of the upcoming \cref{bad-forest-paths-totally-ordered:lem}.
Notice that the cancellation property itself follows straightforwardly from
\cref{equivalent-total-orderings:lem}.

\begin{lemma}[restate={cancellation:lem},name={Cancellation Property}]
    \label{cancellation:lem}
    Let $M$ be a \kl{totally ordered} finite monoid. Then
    for all $a,b,c \in M$, if $b \jleq a$
    and $abc = ab$ then $bc = b$. Similarly,
    if $cba = ba$ then $cb = b$.
    \proofref{cancellation:lem}
\end{lemma}

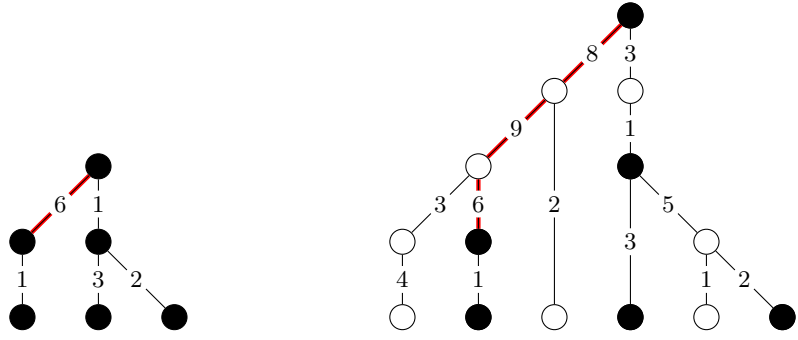
\begin{figure}
    \centering
    \begin{tikzpicture}
        \foreach[count=\i] \x/\y in {0/0,0/1,0/2,1/0,1/1,2/1} {
            \node[draw, circle,fill=black] (t\i) at (\y,\x) {};
        }
        \draw[ultra thick,red] (t4) -- (t6);
        \foreach \i/\j/\k in {4/1/1,5/2/3,5/3/2,6/4/6,6/5/1} {
            \draw (t\j) --
                node[midway,fill=white,circle,inner sep=1pt] {\small $\k$}
                (t\i);
        }
        \foreach[count=\i] \x/\y in {0/0,0/1,0/2,0/3,0/4,0/5,1/0,1/1,1/4,2/1,2/3,3/2,3/3,4/3} {
            \node[draw, circle] (h\i) at (\y+5,\x) {};
        }

        \draw[ultra thick,red] (h8) -- (h10) 
                              -- (h12)
                              -- (h14);

        \foreach \i/\j/\k in {7/1/4,8/2/1,9/5/1,9/6/2,10/7/3,10/8/6,11/4/3,11/9/5,12/10/9,12/3/2,13/11/1,14/12/8,14/13/3} {
            \draw (h\j) --
                node[midway,fill=white,circle,inner sep=1pt] {\small $\k$}
                (h\i);
        }
        \foreach \j in {2,4,6,8,11,14} {
            \node[draw, circle, fill=black] at (h\j) {};
        }

    \end{tikzpicture}
    \caption{A gap embedding between two trees, represented
        using black nodes. The path from the root
        to the left node labelled with the number $6$ is turned
        into a path from the root to a node labelled by
        numbers $8,9,6$. Note that the last number in the path
        is the same as the original label of the edge, and that
        the other numbers are greater or equal to the original label.}
    \label{gap-embedding:fig}
\end{figure}

\AP We are now ready to state the main result of this section relating the
\kl{tree-model embedding} relation to the \kl{gap embedding} relation, in the
case of \kl{totally ordered} monoids. Let $(Q, \leq)$ be an ordered set of
vertex labels, and $(L, \leq)$ be an ordered set of edge labels. Given two
trees $T$ and $T'$, a \intro{gap embedding} from $T$ to $T'$ is a map $h \colon
V(T) \to V(T')$ that respects the ancestor relation, respects the labels on the
vertices, and such that for all edge $x \leq y$ in the input tree, the
corresponding \emph{path} $h(x) \leq h(y)$ in the output tree is labelled
solely with edge labels that are \emph{greater or equal} than the edge label of
$x \leq y$, and such that the last edge of the path \emph{equals} the edge
label of $x \leq y$. We write $T \intro*\gapleq  T'$ if there exists a \kl{gap
embedding} from $T$ to $T'$. To illustrate the notion of \kl{gap embedding}, we
provide an example in
\cref{gap-embedding:fig}.

\begin{theorem}[restate={gap-embedding-wqo:thm},name={\cite{DERSHOWITZ200380}}]
    \label{gap-embedding-wqo:thm}
    Let $(Q, \leq)$ be a \kl{well-quasi-ordered} set of vertex labels,
    and $(L, \leq)$ be a \emph{finite} set of totally ordered edge labels.
    The collection of labelled trees equipped with the \kl{gap embedding relation}
    is \kl{well-quasi-ordered}.
    \proofref{gap-embedding-wqo:thm}
\end{theorem}

\AP In order to recover the \kl{tree-model embedding} relation from the \kl{gap
embedding} relation, we will crucially use the cancellation property
(\cref{cancellation:lem}). To be able to apply such cancellations, we need to
ensure that some elements will be correctly ordered, which is the purpose of
the following definition.

\begin{definition}[Layered intepretation]
    Let $M$ be a \kl{totally ordered} finite monoid.
    We define the \intro{layered interpretation}
    $\intro*\layered \colon M^* \to M^{M}$ as the morphism:
    \begin{equation*}
        (\reintro*\layered(w))_a
        \defined
        \prod_{i=1}^{|w|} \begin{cases} w_i & \text{ if } a \jleq w_i \\ 1_M & \text{ otherwise} \end{cases}
        \quad .
    \end{equation*}
\end{definition}

\begin{theorem}
    \label{gap-embedding-tree-model:thm}

    Let $M$ be a \kl{totally ordered} finite monoid. Then the function
    $\intro*\mtogap$ that maps a \kl{tree-model} $\aTreeModel$ to the tree
    obtained adding as labels to the vertices the value
    $\layered(\lambda(x,r))$ from the vertex to the root of the tree model,
    equipped with the \kl{gap embedding relation} arising from $\jleq$, is an
    \kl{order reflection}.

    In particular, the \kl{tree-models} are \kl{$\infty$-well-quasi-ordered}
    for the \kl{tree-model embedding}.

\end{theorem}
\begin{proof}

    Let $T_1$ and $T_2$ be two \kl{tree models}. Assume that $\mtogap(T_1)
    \gapleq \mtogap(T_2)$. Our goal is to conclude that $T_1 \tmleq T_2$. For
    that, we will simply consider the map $h \colon T_1 \to T_2$ that underlies
    the \kl{gap embedding} relation from $\mtogap(T_1)$ to $\mtogap(T_2)$,
    since the two trees have the same underlying sets of nodes. By definition,
    it respects the ancestor relation, and the labels on the nodes. Let $x,y$
    be two nodes in $T_1$ such that $y$ is the parent of $x$, and let $m$ be
    the label of the edge $x \leq y$ in $T_1$.

    By definition, we know that the path $p$ from $h(x)$ to $h(y)$ is labelled
    with values $u \in M$ that are greater or equal to $m$ for the $\jleq$
    relation. Note that $h(r)$ is the root of the tree $T_2$. Furthermore, we
    know that the label of $x$ is $\layered(\lambda(x,r))$, and the label of
    $h(x)$ is also $\layered(\lambda(h(x),h(r)))$ by definition of $\mtogap$.
    Because $h$ respects the label, we conclude to the following
    equality:
    \begin{equation*}
        \layered(\lambda(x,r)) = \layered(\lambda(h(x),h(r))) \quad .
    \end{equation*}

    Note that the path from the root of the tree $T_1$ to $x$ is labelled with
    $u m$ where $u \in M$, because the last edge ($x \leq y$) is labelled with
    $m$. Similarly, the path from the root of the tree $T_2$ to $h(x)$ is
    labelled with $v w m$, where $w$ is a product of monoid elements that are
    greater or equal to $m$ for the $\jleq$ relation. As a consequence,
    $\layered(\lambda(x,r))_m = \layered(u m)_m = \layered(u)_m m = \layered(v
    w m)_m = \layered(v)_m w m$. Notice that we also have that $\layered(u)_m =
    \layered(v)_m$ because of the labels on $y$ and $h(y)$ are the same. Now,
    using the cancellation property of \cref{cancellation:lem}, we conclude
    that $m = w m$, which is precisely stating that the label of the edge $x
    \leq y$ in $T_1$ is the same as the label of the path $h(x) \leq h(y)$ in
    $T_2$. We have shown that $T_1 \tmleq T_2$.
\end{proof}

\begin{remark}[restate={gap-embedding-tree-model-remark:rem},name={Gap Embedding and Tree Model Embedding}]
    \label{gap-embedding-tree-model-remark:rem}
    It is also quite easy to encode the \kl{gap embedding} relation in terms of
    \kl{tree model embedding}, using a variation 
    of the monoid $M \defined (\set{1, \dots, m}, \min)$.
    \proofref{gap-embedding-tree-model-remark:rem}
\end{remark}

\AP We recover the difficult implication of \cite[Theorem 3]{DRT10} as a
corollary of \cref{gap-embedding-tree-model:thm}.

\subsection{Relationship with Totally Ordered Monoids}
\label{totally-ordered-linear-clique-width:sec}

Let us now relate our characterization of classes of \kl{bounded linear
clique-width} that are \kl{$\infty$-well-quasi-ordered} to the one introduced
by \cite{DRT10}. In particular, we will prove the following theorem, that
witnesses the collapse of \kl{$k$-well-quasi-ordered} to $2$-well-quasi-ordered
when the \kl{edge selector} $P_{\text{edge}}$ is not taken into account.
The main contribution is \cref{bad-forest-paths-totally-ordered:lem} that
precisely connects the cancellation properties of \kl{totally ordered monoids}
(\cref{cancellation:lem}) to the combinatorial properties of \kl{bad forest
paths}. From this, \cref{good-specialisation:thm} follows immediately.

\begin{lemma}[restate={bad-forest-paths-totally-ordered:lem},name={Bad Forest Paths in Totally Ordered Monoids}]
    \label{bad-forest-paths-totally-ordered:lem}
    Let $M$ be a \kl{totally ordered} finite monoid
    and $P_{\text{edge}} \subseteq M^3$ be an \kl{edge selector}.
    Then, there are no \kl{bad forest paths} of length
    greater than $2$
    in $\RegMG{d}$ for all $d \in \Nat$.
    \proofref{bad-forest-paths-totally-ordered:lem}
\end{lemma}

\begin{theorem}[restate={good-specialisation:thm},name={Specialisation of the Main Theorem}]
    \label{good-specialisation:thm}
    Let $M$ be a finite monoid. 
    The following are equivalent:
    \begin{enumerate}
        \item \label{good-spec-totord:item}
            The monoid $M$ is \kl{totally ordered},
        \item \label{good-spec-lwqo:item}
            For all \kl{edge selectors} $P_{\text{edge}} \subseteq M^3$, for all $d \in \Nat$,
            the class of graphs obtained by \kl{monoid downcasting} of $\RegMG{d}$
            is \kl{$\infty$-well-quasi-ordered},
        \item \label{good-spec-2wqo:item}
            For all \kl{edge selectors} $P_{\text{edge}} \subseteq M^3$, for all $d \in \Nat$,
            the class of graphs obtained by \kl{monoid downcasting} of $\RegMG{d}$
            is \kl{$2$-well-quasi-ordered}.
    \end{enumerate}
\end{theorem}
\begin{proof}

    The implication \cref{good-spec-totord:item} $\Rightarrow$
    \cref{good-spec-lwqo:item} follows from
    \cref{bad-forest-paths-totally-ordered:lem} and \cref{main-theorem:thm}.
    Then, the implication \cref{good-spec-lwqo:item} $\Rightarrow$
    \cref{good-spec-2wqo:item} always holds regardless of the class considered.

    Let us prove \cref{good-spec-2wqo:item} $\Rightarrow$
    \cref{good-spec-totord:item} by contraposition. Assume that $M$ is not
    \kl{totally ordered}, then there exists $a,b \in M$ such that $\jideal{ab}
    \neq \jideal{a}$ and $\jideal{ab} \neq \jideal{b}$. Let us write
    $\aMgraph[A]$ to be the \kl{monoid-labelled graph} evaluating to $a$ with a
    single vertex $v$ having right label $1_M$ and left label $a$. Similarly,
    let us define $\aMgraph[B]$ to be the \kl{monoid-labelled graph} evaluating
    to $b$ with a single vertex $v$ having right label $1_M$ and left label
    $b$. Using the \kl{edge selector} $P_{\text{edge}} = \setof{(x,y,z)}{x \in
    M, z \in M, y \in \set{a,b}}$, we construct a sequence of graphs
    $\aMgraph_0 \defined \aMgraph[A] \bp \aMgraph[B]$, and $\aMgraph_{n+1}
    \defined \aMgraph_0 \bp \aMgraph_{n}$. Remark that by construction, the
    sequence $\aMgraph_n$ contains arbitrarily long induced paths. As a
    consequence, this sequence of graphs is not \kl{$2$-well-quasi-ordered}.
    Thanks to \cref{factorization-forest-theorem:thm},
    these graphs all belong to $\RegMG{d}$, for a fixed $d$.
    Hence, the class of graphs obtained by \kl{monoid downcasting} of $\RegMG{d}$
    is not \kl{$2$-well-quasi-ordered}.
\end{proof}
\section{Outlook}
\label{conclusion:sec}

\subparagraph{The Pouzet Conjecture.} We have provided in the case of classes
of \kl{bounded linear clique-width} a partial answer to the Pouzet conjecture
by showing that \cref{pouzet2:conj} holds, and that a parametrized version of
\cref{pouzet1:conj} holds. There are two main directions to pursue in our
opinion. The first one is to try to lower the bound $k$ in
\cref{main-theorem:thm} to be $2$, by a suitable analysis of the \kl{bad forest
paths} and their combinatorics. This could also benefit from the ability to
construct so-called \emph{optimal} decompositions of graphs \cite{BOPIL17},
that are more likely to exhibit bad patterns. A second direction is to try to
understand whether the results can be adapted to tree-like structures, and in
particular to \kl{bounded clique-width}. This would probably require the use of
analogues of the factorization forest theorem in the tree setting such as
\cite{COLC07}.

\subparagraph{Gap Embeddings.} We have proven that there is a connection
between the \kl{gap embedding} relation and the fact that some classes of
graphs are \kl{well-quasi-ordered} by the induced subgraph relation. This can
be seen in \cref{gap-embedding-tree-model:thm}, but is actually already present
in a subtle form in \cref{main-theorem:thm}. Indeed, the \kl{gap embedding}
relation on words (linear trees) can be understood as a nested \kl{subword
embedding} relation, which is how \cref{main-theorem:thm} is proven (by nesting
several times the \kl{subword embedding} relation). This suggests that to work
on the Pouzet conjectures for classes of graphs of \kl{bounded clique-width},
one should try to understand a form of nested \kl{Kruskal embeddings}, which
characterizes the \kl{gap embedding} in general \cite{FREU20}.

\subparagraph{Labelled Well-Quasi-Orderings.} We conjecture that classes of
graphs that are \kl{$\infty$-well-quasi-ordered} are all of \kl{bounded
clique-width}, which is a weakening of the conjecture that hereditary classes
that are \kl{$2$-well-quasi-ordered} have \kl{bounded clique-width}
\cite[Conjecture 5, p. 13]{DRT10}. This could provide a positive answer to the
second Pouzet conjecture (\cref{pouzet2:conj}) in a two-step fashion, which
would be a significant contribution to the field. A first possibility would be
to restrict our attention to \emph{dependent classes} of graphs, where the
recent advances in their structural theory could be of great help
\cite{DREI23,OHPT23}. While this does not close the Pouzet conjecture
(\cref{pouzet1:conj}), even in the case of \kl{bounded linear clique-width}, it
provides a new perspective on the problem, and tools to tackle it.

\subparagraph{Finite Power Property.} The questions answered in this paper are
connected to research on well-quasi-orders that can be defined on $\Sigma^*$ by
sets of rewrite rules since a \kl{tree-model embedding} is essentially obtained
by rewriting an element $m$ of a finite monoid $M$ by a product $m_1 \cdots
m_n$ of elements of $M$ evaluating to $m$. Such rewriting systems have been
studied in the past and the main results are that the \emph{unavoidability} of
some patterns prevents the existence of large antichains
\cite{EHR83,BEH85,KUNC05}. Pursuing this line of research, it was proven that
\kl{totally ordered monoids} are precisely those that recognize solely
languages enjoying the \intro{finite power property}, i.e., languages such that
$\bigcup_{n \geq 0} L^n$ stabilizes after finitely many steps. We conclude by
the following intriguing remark: \cite{KUNC05,KIRS02}. It remains unclear
whether there is a direct proof of the result of \cite{DRT10} using the finite
power property, but it is aesthetically pleasing that, at least intuitively,
such a property precisely forbids the existence of ``long induced paths'' and
their variants (long induced complement of paths, etc.).

\bibliographystyle{plainurl}
\section{Proofs of \cref{preliminaries:sec}}

\begin{proofof}{hereditary-closure:lemma}
    Let $\Cls$ be a class of graphs and $k$ be such 
    such that $\Label{k+1}(\Cls)$ is \kl{well-quasi-ordered}.
    Let us consider $\Cls[D]$ be defined as the \kl{hereditary closure} of $\Cls$.
    We have to prove that $\Label{k}(\Cls[D])$ is \kl{well-quasi-ordered}.

    Consider a sequence $\seqof{G_i}{i \in \Nat}$ of graphs in
    $\Label{k}(\Cls[D])$. By definition, there exists a sequence of graphs
    $\seqof{H_i}{i \in \Nat}$ in $\Cls$ such that $G_i$ is an \kl{induced
    subgraph} of $H_i$. We can label $H_i$ with $k + 1$ labels as follows: we
    place an ``unused" label on the vertices of $H_i$ that do not belong to
    $G_i$, and we use the $k$ labels of $G_i$ for the vertices of $G_i$.

    Because $\Label{k + 1}(\Cls)$ is \kl{well-quasi-ordered}, the constructed
    sequence is is \kl(sequence){good}, and there exists $i < j$ such that $H_i
    \isubleq H_j$ respecting the labels. Note that this immediately implies
    $G_i \isubleq G_j$.
\end{proofof}

\begin{proofof}{bd-lcw-2-wqo:thm}
    Let $\Cls$ be a class of graphs that has \kl{bounded linear clique-width}
    and is \kl{$2$-well-quasi-ordered}.
    Because it has \kl{bounded linear clique-width}, $\Cls$ is included in the image of a
    $\MSO$-interpretation $I_0$ from finite words to graphs. Furthermore,
    because $\Cls$ is \kl{$2$-well-quasi-ordered}, there are finitely many
    minimal obstructions for $\Cls$ inside $I_0(\Sigma^*)$ \cite[Proposition
    3]{DRT10}. Because the set of obstructions is finite, it can be encoded in
    the formula $\varphi_{\Delta}$. This ensures that the image of the new
    interpretation is contained in the \kl{hereditary closure} of $\Cls$ as
    needed.
\end{proofof}

\begin{proofof}{bd-lcw-2-wqo:cor}
    Assume that the second Pouzet conjecture (\cref{pouzet2:conj})
    holds for all images of \kl{$\MSO$-interpretations}.

    Let $\Cls$ be a class of graphs that has \kl{bounded linear clique-width},
    and is \kl{$\infty$-well-quasi-ordered}. By \cref{bd-lcw-2-wqo:thm}, there
    exists an $\MSO$-interpretation $I$ such that $\Cls \subseteq \msoIm(I)
    \subseteq \dwclosure[\isubleq]{\Cls}$. Since $\Cls$ is
    \kl{$\infty$-well-quasi-ordered}, so is $\dwclosure[\isubleq]{\Cls}$ by
    \cref{hereditary-closure:lemma}. As a consequence, $\msoIm(I)$ is also
    \kl{$\infty$-well-quasi-ordered}. By assumption, it means that $\msoIm(I)$
    is \kl{labelled-well-quasi-ordered}, and then so is $\Cls$ as a subset.

\end{proofof}

\section{Proofs of \cref{bounded-lcw:sec}}

\begin{proofof}{removing-delta-dom:lemma}
    To remove $\varphi_{\Delta}$, remark that we can write
    $\varphi'_{\text{dom}}(x) = \varphi_{\text{dom}}(x) \land
    \varphi_{\Delta}$. The image is therefore the empty graph $G_{\emptyset}$
    whenever $\varphi_{\Delta}$ does not hold, and we obtain that $\msoIm(I) =
    \msoIm(I') \cup \set{ G_\emptyset }$.

    To remove $\varphi_{\text{dom}}$, we can write $\varphi'_{\text{edge}}(x,y)
    = \varphi_{\text{edge}}(x,y) \land \varphi_{\text{dom}}(x) \land
    \varphi_{\text{dom}}(y)$. As a consequence, the graphs in the
    \kl(interpretation){image} of $I'$ are exactly the graphs of $I$ plus an
    independent set. Now, because the class of independent sets is
    \kl{labelled-well-quasi-ordered} by the \kl{induced subgraph} relation, the
    result follows.
\end{proofof}

\begin{proofof}{bad-forests-paths-imply-antichains:lemma}

    Assume that we have access to arbitrarily long \kl{bad forest paths}. Let
    $\seqof{\aMgraph_i}{i \in \Nat}$ be a sequence of \kl{monoid-labelled
    graphs} where $\aMgraph_i \defined \ep{\seqof{\aMgraph[F]_i^j}{1 \leq j
    \leq n_i}}$ is a \kl{bad forest path} of length $n_i$. We can extract from
    this sequence to ensure that $n_i$ is strictly increasing. Furthermore, we
    can extract so that $n_{i+1}$ is greater than the number of vertices in
    $\aMgraph_i$.

    Now, because these are \kl{bad forest paths}, for all $i \in \Nat$, there
    exists $a,b \in M^2$ such that $\mcast{a \aMgraph_i b}$ cannot be embedded
    into a long product. By extracting further,  we can assume that $a$ and $b$
    are the same for all the \kl{monoid-labelled graphs} in the sequence.

    Let us now consider the sequence of graphs $\seqof{G_i}{i \in \Nat}$ where
    $G_i \defined \mcast{a \aMgraph_i b}$, and let us place labels on the
    vertices of $G_i$ as follows. First, on every vertex, we place as label the
    monoid element that labelled it in $\aMgraph_i$. Then, we also place a
    label on the vertices that come from $\aMgraph[F]_i^1$ to distinguish the
    ``starting vertices", a label on the vertices that come from
    $\aMgraph[F]_i^{n_i}$ to distinguish the ``ending vertices". Formally, the
    labels are taken from $M^2 \times \set{ \text{start}, \text{middle},
    \text{end}}$, which is of size $|M|^2 \times 3$.

    Assume by contradiction that there exists an embedding $h$ between two
    graphs $G_i$ and $G_j$ that respects the labels. Since $i < j$, $n_j$ is
    greater than the number of vertices in $G_i$, therefore there exists $1
    \leq \ell \leq n_j$ such that $\aMgraph[F]_j^\ell$ is not in the image of
    $h$. The embedding $h$ proves that $\seqof{\aMgraph[F]_i^{\ell}}{1 \leq
    \ell \leq n_i}$ is a \kl{good forest path} (with context $a,b$),
    which is absurd.

    We have proven that 
    $\set{ \mcast{a \aMgraph_i b} }_{i \in \Nat}$ is not
    \kl{$k$-well-quasi-ordered}, with $k = 3 \times |M|^2$.
\end{proofof}

\begin{figure}
    \centering
    \begin{tikzpicture}
        \draw (0,0) rectangle (4,1);
        \draw (1,0) -- (1,1);
        \draw (2,0) -- (2,1);
        \draw (3,0) -- (3,1);
        \node at (0.5,-0.5) {$\aMgraph_1$};
        \node at (1.5,-0.5) {$\aMgraph_2$};
        \node at (2.5,-0.5) {$\aMgraph_3$};
        \node at (3.5,-0.5) {$\aMgraph_4$};
        \node at (0.5, 1.3) {$e$};
        \node at (1.5, 1.3) {$e$};
        \node at (2.5, 1.3) {$e$};
        \node at (3.5, 1.3) {$e$};
        \draw (0,1.1) -- (4,1.1);
        \draw[fill=C1] (0,0) rectangle (1,1);
        \draw[fill=C3] (3,0) rectangle (4,1);
        \draw[fill] (1.5,0.3) circle (0.05);
        \draw[fill] (1.5,0.7) circle (0.05);
        \draw[fill] (2.5,0.3) circle (0.05);
        \draw[fill] (2.5,0.7) circle (0.05);

        \begin{scope}[yshift=-3cm,xshift=-2cm]
            \draw (0,0) rectangle (8,1);
            \node at (0.5,1.3) {$e$};
            \node at (7.5,-0.5) {$\aMgraph[H]_8$};
            \foreach \i in {1,...,7} {
                \draw (\i,0) -- (\i,1);
                \node at (\i + 0.5, 1.3) {$e$};
                \node at (\i - 0.5, -0.5) {$\aMgraph[H]_{\i}$};
            }
            \draw (0,1.1) -- (8,1.1);
            \draw[fill=C2] (4,0) rectangle (5,1);
            \draw[fill=C1] (0,0) rectangle (1,1);
            \draw[fill=C3] (7,0) rectangle (8,1);
            \coordinate (P1) at (1.5, 0.5);
            \coordinate (P2) at (2.5, 0.5);
            \coordinate (P3) at (6.5, 0.5);
            \coordinate (P4) at (5.5, 0.5);

            \draw[fill] (P1) circle (0.05);
            \draw[fill] (P2) circle (0.05);
            \draw[fill] (P3) circle (0.05);
            \draw[fill] (P4) circle (0.05);
        \end{scope}

        \draw[dashed, ->] (1.5,0.3) -- (P1);
        \draw[dashed, ->] (1.5,0.7) -- (P3);
        \draw[dashed, ->] (2.5,0.3) -- (P2);
        \draw[dashed, ->] (2.5,0.7) -- (P4);
    \end{tikzpicture}
    \caption{Illustrating the proof of \cref{good-forest-paths-decomp:lemma}.
        (\cref{good-forest-paths-decomp:lemma:proof}
         page \pageref{good-forest-paths-decomp:lemma:proof}).
         The original \kl{good forest path}
         described on the top of the picture is embedded
         into a larger sequence on the bottom. The endpoints
         are distinguished by colors, and the
         red distinguished graph $\aMgraph[H]_5$ is untouched by
         the embedding. We illustrate in dashed lines some
         of the mappings of the embedding from the top sequence to
         the bottom one.}
    \label{good-forest-paths-decomp:fig}
\end{figure}
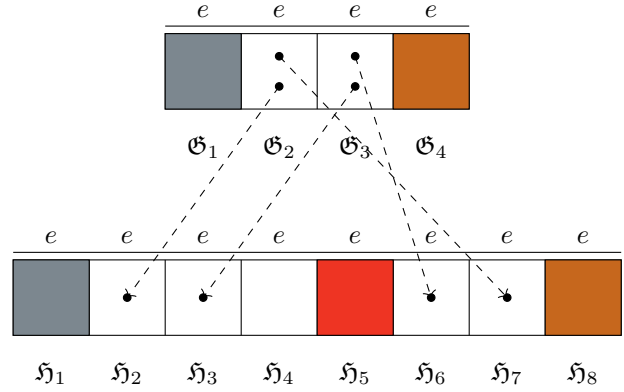

\begin{proofof}{good-forest-paths-decomp:lemma}

    The proof is relatively simple. Assume that $\seqof{\aMgraph_i}{1 \leq i
    \leq n}$ is a \kl{good forest path}. Let us consider a context $a,b \in
    M^2$. By definition, there exists another sequence $\seqof{\aMgraph[H]_i}{1
    \leq i \leq m}$ together with an \kl(graph){embedding} from $G$ to $H$,
    where $G \defined \mcast{a \ep{\seqof{\aMgraph_i}{1 \leq i \leq n}}
    b}$ and $H \defined \mcast{a \ep{\seqof{\aMgraph[H]_i}{1 \leq i \leq
    m}} b}$, such that $h$ preserves the labels of the vertices, the image of
    $\aMgraph_1$ is contained in $\aMgraph[H]_1$, the image of $\aMgraph_n$ is
    contained in $\aMgraph[H]_m$, and there exists an $1 \leq i \leq m$ such
    that the image of $h$ does not intersect $\aMgraph[H]_i$. This is depicted
    in \cref{good-forest-paths-decomp:fig} in the specific case where $a = b =
    1_M$.

    Now, let us embed the sequence $\seqof{\aMgraph_i}{1 \leq i \leq n}$ into
    the same sequence repeated three times. Let us write those graphs
    $\seqof{\aMgraph[C]_{i,j}}{1 \leq i \leq n, 1 \leq j \leq 3}$, where
    $\aMgraph[C]_{i,j} \defined \aMgraph_i$. Our embedding $g$ is defined as
    follows: $g$ sends a note $x \in \aMgraph_i$ to the same node $x$ in
    $\aMgraph[C]_{i,j}$ where $j = 1$ or $j = 3$, depending on whether $h(x)$
    is sent to a node that is \emph{on the left}  of the untouched graph
    $\aMgraph[H]_i$ ($j = 1$) or \emph{on the right} of the untouched graph
    $\aMgraph[H]_i$ ($j = 3$).

    Note that in particular, $g$ is the identity on the first and last graphs
    $\aMgraph_1$ and $\aMgraph_n$, and sends them to the first and last graphs
    of our new sequence $\aMgraph[C]_{1,1}$ and $\aMgraph[C]_{n,3}$.
    Furthermore, by definition, it respects the labelling. Finally, by
    construction, no graphs of the form $\aMgraph[C]_{i,2}$ are in the image of
    $g$. It only remains to check that our function $g$ is an
    \kl(graph){embedding} from $G$ to $C \defined \mcast{a
    \ep{\seqof{\aMgraph[C]_{i,j}}{1 \leq i \leq n, 1 \leq j \leq 3}} b}$.

    However, it is immediate from the fact that $h$ was an
    \kl(graph){embedding} that $g$ is: an edge between $x$ and $y$ in $G$
    exists if and only if there is an edge between $h(x)$ and $h(y)$ in $H$,
    which only depends on the label of $h(x)$, $h(y)$, their relative ordering
    and whether they are in consecutive graphs of the sequence
    $\seqof{\aMgraph[H]_i}{1 \leq i \leq m}$.
    Let us now consider $x$ and $y$. If they are both sent to the same
    copy of the sequence ($j = 1$ or $j = 3$), then
    their behaviour is exactly the same as in the graph $G$
    hence there is an edge in $C$ if and only if there is an edge in $G$.

    If they are separated, their behaviour is the same as the one of $h(x)$ and
    $h(y)$: they have the same labels, the same relative ordering, and are both
    separated by idempotent graphs. Hence,
    there is an edge between $h(x)$ and $h(y)$ if and only if
    there is an edge between $x$ and $y$. Because $h$ was an \kl(graph){embedding},
    this allows us to conclude.
\end{proofof}

\begin{proofof}{bounded-forest-paths:theorem}
    One implication follows directly from \cref{bad-forests-paths-imply-antichains:lemma}.

    For the converse implication, we by induction on $d$ that the class of
    graphs obtained as the collection of \kl{monoid downcastings} $\mcast{a
    \aMgraph b}$ for all $a, b \in M^2$ and $\aMgraph$ in $\RegMG{d}$ is
    \kl{labelled-well-quasi-ordered}. When $d = 1$, all graphs have size $1$, so the
    image is finite and therefore \kl{labelled-well-quasi-ordered}.

    For $d+1$, we know that $\RegMG{d+1}$ can be split into those graphs
    obtained as the binary product of two graphs in $\RegMG{d}$, and those
    obtained as an \kl{idempotent product} of a sequence of graphs in
    $\RegMG{d}$ all \kl(mlg){evaluating} to the same \kl{idempotent} element $e
    \in M$. To prove that the class of \kl{monoid downcastings} of graphs in
    $\RegMG{d+1}$ is \kl{labelled-well-quasi-ordered}, it suffices to prove
    that these two sets are \kl{labelled-well-quasi-ordered} independently, as
    the union of two \kl{labelled-well-quasi-ordered} sets is
    \kl{labelled-well-quasi-ordered}.

    \textbf{For the binary product.} Let us consider a sequence $\seqof{G_i}{i
    \in \Nat}$ of labelled graphs, where $G_i$ is a labelled version of the
    \kl{monoid downcasting} of $\aMgraph_i \defined \aMgraph[A]_i \bp
    \aMgraph[B]_i$, $\aMgraph[A]_i \in \RegMG{d}$ and $\aMgraph[B]_i \in
    \RegMG{d}$. Let us write $l_i$ for the \kl(mlg){evaluation} of
    $\aMgraph[A]_i$ and $r_i$ for the \kl(mlg){evaluation} of $\aMgraph[B]_i$.
    Let us consider the sequence of pairs $\seqof{(A_i,B_i)}{i \in \Nat}$ where
    $A_i = \mcast{a \aMgraph[A]_i r_i b}$ and $B_i = \mcast{a l_i \aMgraph[B]_i
    b}$. Because $M$ is finite, we can extract the sequence such that $l_i$ and
    $r_i$ are actually constant sequences (do not depend on $i$). Then, by
    induction hypothesis, and thanks to Dickson's lemma, we know that this must
    be a \kl{good sequence} of graphs. As a consequence, there exists $i < j$
    such that $A_i \isubleq A_j$ and $B_i \isubleq B_j$. As a consequence,
    $G_i$ embeds in $G_j$.

    \textbf{For an idempotent product.} Let us consider a sequence
    $\seqof{G_i}{i \in \Nat}$ of labelled graphs such that for all $i \in
    \Nat$, $G_i = \mcast{a \aMgraph_i b}$, with $\aMgraph_i =
    \ep{\seqof{\aMgraph[F]_i^j}{1 \leq j \leq n_i}}$, such that all
    $\aMgraph[F]_i^j$ \kl(mlg){evaluate} to the same \kl{idempotent} element $e
    \in M$ for all $i \in \Nat$ and all $1 \leq j \leq n_i$.

    By assumption, there exists a bound $N$ on the length of \kl{bad forest
    paths}. Let us group the forests in $\seqof{\aMgraph[F]_i^j}{1 \leq i \leq
    n_i}$ in buckets of size $N+1$. By construction all of these buckets are
    \kl{good forest paths}. Now, using \cref{good-forest-paths-decomp:lemma},
    we can decompose each bucket into three copies of itself. This provides us
    with a new sequence of graphs $\seqof{\aMgraph[C]_i^j}{1 \leq j \leq m_i}$
    in which our original sequence embeds, and such that the image of the
    embedding touches at most $2N$ consecutive graphs. Let us give a name to
    these maximal consecutive sequences of graphs that intersect with the
    embedding, by writing $\aMgraph[P]_i^\ell$ for the $\ell$th such product:
    it is a \kl{binary product} of at most $2N$ graphs in the sequence
    $\seqof{\aMgraph[C]_i^j}{1 \leq j \leq m_i}$.

    Leveraging the case of the binary product, we conclude that the collection
    of $\mcast{ae \aMgraph[P]_i^\ell e b}$ for all $i \in \Nat$ and valid
    $\ell$ is \kl{labelled-well-quasi-ordered}. Now, using Higman's lemma
    \cite{HIG52}, we can conclude that there exists $i < j$
    and a strictly increasing function $\rho$ such that
    \begin{equation*}
        \mcast{a e \aMgraph[P]_i^\ell e b} \isubleq \mcast{a e
            \aMgraph[P]_j^{\rho(\ell)} e b}, \quad \forall \ell
            \quad .
    \end{equation*}

    We claim that there exists an embedding from $G_i$ to $G_j$, obtained by
    sending an element $x \in \aMgraph[P]_i^\ell$ into its corresponding
    element in $\aMgraph[P]_j^{\rho(\ell)}$.

    The only edges for which it is unclear whether this definition yields an
    embedding is between two elements $x$ and $y$ in \emph{different} products
    $\aMgraph[P]_i^\ell$ and $\aMgraph[P]_i^{\ell'}$. Note that by
    definition, such products were already separated by an untouched
    idempotent graph. Furthermore, since $\rho$ is strictly increasing,
    the relative ordering of these nodes is preserved. As a consequence,
    the edge relation is preserved by our definition.

    This concludes the proof in the case of an idempotent product.
\end{proofof}

\begin{proofof}{decide-bounded-forest-paths:lemma}

    Let $M$ be a finite monoid and $P_{\text{edge}}$ be an \kl{edge selector}.
    We work over the alphabet $\Sigma \defined M \times M \uplus \set{ \langle
    } \times M \uplus \set { \rangle } \times M$. A word will be of the from
    $\langle_m (a_1, b_1) (a_2,b_2) \cdots \rangle_m \langle_e (c_1, d_1)
    (c_2,d_2) \cdots \rangle_e \cdots$. This is an encoding of a sequence of
    \kl{monoid-labelled graphs}.

    First, we claim that there exists an $\MSO$ formula that checks whether a
    word over $\Sigma$ is \emph{well-formed}: brackets are not nested, the
    product of the pairs of elements inside a bracket evaluate to the element
    decorating the bracket, etc.

    Now, we claim that there exists an $\MSO$ sentence that checks whether such
    a word encodes a \kl{bad forest path}. This is done by leveraging
    \cref{good-forest-paths-decomp:lemma}: if the forest path is good, it
    embeds into a repetition of itself three times by sending nodes to the
    leftmost or rightmost copy of the original sequence. The $\MSO$ sentence
    checks that for all assignments of the nodes of the graphs into a
    \emph{left} or \emph{right} copy, the resulting function from the word to a
    repetition of three times this word is not an \kl(graph){embedding}. 

    Let us now consider the projection of the language
    of well-formed words encoding bad forest paths onto the alphabet
    $M \times \set{ \langle, \rangle }$. This remains a regular language,
    and there are unbounded bad forests paths if and only if this language 
    is unbounded, which is decidable.
\end{proofof}

\begin{proofof}{gap-embedding-tree-model-remark:rem}
    We consider a first monoid $M_1 \defined (\set{1, \dots, m}, \min)$, and
    a second monoid $M_2 \defined
    (\set{ 1, \dots, m}, \lambda xy. y)$. The monoid $M \defined M_1 \times
    M_2$ is \kl{totally ordered}, because $(a,b)(c,d) = (\min(a,c), d)$, hence
    if $a \leq c$, $\jideal{(a,b)(c,d)}) = \jideal{(c,d)}$, and otherwise
    $\jideal{(a,b)(c,d)}) = \jideal{(a,d)} = \jideal{(a,b)}$ by multiplication
    by $(a,b)$ on the right.

    Furthermore, the \kl{tree-model embedding} relation from $T_1$ to $T_2$ is
    precisely stating that the labels inserted must be greater or equal on the
    first component, and that the last edge must have the same label on the
    second component. If we restrict our attention to trees where labels are of
    the form $(a,a)$, this precisely encodes the \kl{gap embedding} relation.
\end{proofof}

\section{Proofs of \cref{totally-ordered-monoids:sec}}

\begin{lemma}
    \label{idempotent-groups:lem}
    Let $M$ be a finite monoid such that
    for all $x,y$ either $\rideal{x} \subseteq \rideal{xy}$
    or $\rideal{y} \subseteq \rideal{xy}$.
    Then, there exists $k \geq 2$ such that
    for all $x \in M$,
    $x^k = x$.
\end{lemma}
\begin{proof}
    Let $x \in M$, and remark that the property of $M$ implies $\rideal{x}
    \subseteq \rideal{xx}$, which tells us that $\rideal{x} = \rideal{xx}$, and
    by induction $\rideal{x} = \rideal{x^n}$ for all $n \in \Nat$. Consider $n$
    such that $x^n$ is \kl{idempotent} which exists because $M$ is finite.
    Notice that there exists $u \in M$ such that $x^n u = x$, because
    $\rideal{x^n} = \rideal{x}$. However, $x^n x^n = x^n$ because it is an
    \kl{idempotent}. As a consequence, $x^n x^n u = x^n u = x$, and $x^n = x$.
    Because $M$ is finite, one can actually find a unique $k$ such that
    $x^k = x$ for all $x \in M$.
\end{proof}

\begin{lemma}
    \label{pseudo-inverse:lem}
    Let $M$ be a finite monoid such that
    for all $x,y$ either $\rideal{x} \subseteq \rideal{xy}$
    or $\rideal{y} \subseteq \rideal{xy}$.
    Then, for all $x,y \in M^2$,
    there exists $k \geq 1$
    such that $(xy)^k x = x$ or $(yx)^k y = y$.
\end{lemma}
\begin{proof}
    Let $x,y \in M$. We know that either $\rideal{x} \subseteq \rideal{xy}$ or
    $\rideal{y} \subseteq \rideal{xy}$. Without loss of generality, let us
    assume that $\rideal{x} \subseteq \rideal{xy}$. Using
    \cref{idempotent-groups:lem}, we conclude that $(xy)^k = xy$ for some $k >
    1$. Because $\rideal{x} \subseteq \rideal{xy}$, there exists $u \in M$,
    such that $xyu = x$. Multiplying the previous equation by $u$, we obtain
    $(xy)^{k-1} xyu = xyu = x$. This proves that 
    $(xy)^{k-1} x = x$ as expected.
\end{proof}

\begin{lemma}
    \label{jideal-implies-rideal:lem}
    Let $M$ be a finite monoid 
    and $x,y \in M$ be such that $\jideal{xy} = \jideal{x}$.
    Then $\rideal{x} \subseteq \rideal{xy}$.
\end{lemma}
\begin{proof}
    By definition, there exists $u,v \in M$,
    such that $uxyv = x$.
    Using an immediate induction,
    for all $k \in \Nat$,
    $u^k x (yv)^k = x$.
    Because $M$ is finite, there exists $k \geq 2$
    such that $(yv)^k$ is \kl{idempotent}.
    Then,
    \begin{equation*}
        x = u^k x (yv)^k = u^k x (yv)^k (yv)^k = x (yv)^k
        \quad .
    \end{equation*}
    We conclude that $\rideal{x} \subseteq \rideal{xy}$.
\end{proof}

\begin{proofof}{equivalent-total-orderings:lem}

    First of all, if $\rideal{x} \subseteq \rideal{xy}$, then $\jideal{xy} =
    \jideal{x}$, and similarly for \kl{left ideals}. As a consequence,
    \cref{equivalent-total-orderings-rideal:item} and
    \cref{equivalent-total-orderings-lideal:item} imply
    \cref{equivalent-total-orderings-jideal:item}.

    Then, it is quite immediate that
    \cref{equivalent-total-orderings-jideal:item} is equivalent to
    \cref{equivalent-total-orderings-jideal2:item}. Indeed, we always have
    $\jideal{xy} \subseteq \jideal{x}$ and $\jideal{xy} \subseteq \jideal{y}$.
    As a consequence, if $\jideal{xy} = \jideal{x}$, then $\jideal{xy} =
    \jideal{x} \cap \jideal{y}$, and therefore $\jideal{x} \subseteq
    \jideal{y}$. The converse is immediate.

    The only non-trivial implication is to prove that
    \cref{equivalent-total-orderings-jideal:item} implies
    \cref{equivalent-total-orderings-rideal:item}. Then, by considering the
    dual monoid $M^{\text{op}}$, we will also conclude that
    \cref{equivalent-total-orderings-jideal:item} implies
    \cref{equivalent-total-orderings-lideal:item}.
    However, this is exactly the content of \cref{jideal-implies-rideal:lem}.

    Let us conclude the proof by considering a finite set $Q$ and a monoid $M$
    of endofunctions of $Q$ endowed with function composition and showing that
    the property of \cref{equivalent-total-orderings-jideal:item} is equivalent
    to being totally ordered for the relation $f \dalileq g \defined \image{f}
    \subseteq \image{f \circ g}$.

    Remark that if $\rideal{f} \subseteq \rideal{f \circ g}$, there exists $h
    \in M$ such that $f = f \circ g \circ h$ and therefore $f \dalileq g$.
    Hence, if $M$ satisfies \cref{equivalent-total-orderings-jideal:item}, then
    it is totally ordered for $\dalileq$.

    Conversely, assume that $M$ is totally ordered for $\dalileq$. One can
    prove statements analogue to the ones of
    \cref{jideal-implies-rideal:lem,pseudo-inverse:lem,idempotent-groups:lem}
    for the relation $\dalileq$.
    The key argument is to notice that if $f \dalileq g$, then there exists $h
    \colon Q \to Q$ that \emph{may not be in $M$}, such that $f = f \circ g
    \circ h$, but that in the proofs this particular $h$ \emph{vanishes}. Let us 
    illustrate this fact for the particular case of \cref{jideal-implies-rideal:lem}.
    Let $k \geq 2$ be such that $(f \circ g)^k$ is an \kl{idempotent} of $M$.
    Notice that $f \dalileq (f \circ g)^k$ by induction. As a consequence,
    there exists $h \colon Q \to Q$, such that $f = (f \circ g)^k \circ h$, but
    since $(f \circ g)^k$ is \kl{idempotent}, we conclude that $f = (f \circ
    g)^k \circ f$. In particular, we have proven that $\rideal{f} \subseteq
    \rideal{f \circ g}$.
\end{proofof}

\begin{proofof}{cancellation:lem}
    Let $a,b,c \in M$ be such that $\jideal{ab} = \jideal{b}$,
    and $abc = ab$.
    Then, $\lideal{b} \subseteq \lideal{ab}$ thanks to
    \cref{equivalent-total-orderings:lem}.
    In particular, there exists $u \in M$ such that $uab = b$.
    Multiplying by $u$ on the left the equation
    $abc = ab$, we conclude that $bc = c$ as expected.

    For the similar equation $cba = ba$, we 
    use the \kl{right ideals} instead and conclude similarly.
\end{proofof}

\begin{proofof}{gap-embedding-wqo:thm}
    It is known from \cite{DERSHOWITZ200380} that the \kl{gap embedding relation}
    is a \kl{well-quasi-ordering}.
    However, we have a slightly different notion of \kl{gap embedding} than the
    authors, because we enforce that the label of the last edge of the path
    is kept equal to the label of the edge in the original tree.

    Let us briefly sketch why this is not a problem. Given a collection $(Q,
    \leq)$ of labels on the vertices, and $(L, \leq)$ on the edges, we
    construct the set $(Q, \leq) \times (\set{ 1 } + L, =)$ of labels ordered
    pointwise on the vertices. This new collection of vertex labels is
    \kl{well-quasi-ordered} assuming that $(Q, \leq)$ is
    \kl{well-quasi-ordered} and $L$ is finite. Now, given a tree $T$ we can
    produce a new tree $f(T)$ by replacing the labels of a given vertex $v$ by
    the pair composed of its actual label, and the label of the edge leading to
    its parent if such an edge exists, and $1$ otherwise.

    This provides an \kl{order reflection} from trees
    ordered with the \kl{gap embedding relation}
    into trees ordered with the relation defined by \cite{DERSHOWITZ200380}.
\end{proofof}

\begin{figure}
    \begin{tikzpicture}[
        emb/.style={
            ->,
            >=stealth,
            thick,
            dashed
        }
        ]
        \begin{scope}
            \draw (0,0) rectangle (6,1);
            \draw (2,0) -- (2,1);
            \draw (4,0) -- (4,1);
            \node at (1,-0.5) {$\aMgraph[U]$};
            \node at (3,-0.5) {$\aMgraph[A]$};
            \node at (5,-0.5) {$\aMgraph[V]$};
            \node at (1, 1.3) {$u$};
            \node at (3, 1.3) {$a$};
            \node at (5, 1.3) {$v$};
            \draw (0,1.1) -- (6,1.1);
            \draw[fill] (1,0.5) circle (0.05);
            \draw[fill] (3,0.5) circle (0.05);
            \draw[fill] (5,0.5) circle (0.05);
            \coordinate (OLDX) at (1,0.5);
            \coordinate (OLDY) at (3,0.5);
            \coordinate (OLDZ) at (5,0.5);

            \node at (1,0.2) {$x$};
            \node at (3,0.2) {$y$};
            \node at (5,0.2) {$z$};
            \draw[leftlabel] (0,0.5) --
                node[midway,above,leftlabel] {$u_1$}
                (1,0.5);
            \draw[rightlabel] (1,0.5) --
                node[midway,above,rightlabel] {$u_2$}
                (2,0.5);
            \draw[leftlabel] (2,0.5) -- 
                node[midway,above,leftlabel] {$a_1$}
                (3,0.5);
            \draw[rightlabel] (3,0.5) --
                node[midway,above,rightlabel] {$a_2$}
                (4,0.5);
            \draw[leftlabel] (4,0.5) --
                node[midway,above,leftlabel] {$v_1$}
                (5,0.5);
            \draw[rightlabel] (5,0.5) -- 
                node[midway,above,rightlabel] {$v_2$}
                (6,0.5);
        \end{scope}

        \begin{scope}[yshift=-3cm]
            \draw (0,0) rectangle (6,1);
            \draw (2,0) -- (2,1);
            \draw (4,0) -- (4,1);
            \node at (1,-0.5) {$\aMgraph[U]$};
            \node at (3,-0.5) {$\aMgraph[A]$};
            \node at (5,-0.5) {$\aMgraph[V]$};
            \node at (1, 1.3) {$u$};
            \node at (3, 1.3) {$a$};
            \node at (5, 1.3) {$v$};
            \draw (0,1.1) -- (6,1.1);
            \draw[fill] (1,0.5) circle (0.05);
            \draw[fill] (3,0.5) circle (0.05);
            \coordinate (NEWX) at (1,0.5);
            \coordinate (NEWY) at (3,0.5);

            \node at (1,0.2) {$x$};
            \node at (3,0.2) {$y$};
            \draw[leftlabel] (0,0.5) --
                node[midway,above,leftlabel] {$u_1$}
                (1,0.5);
            \draw[rightlabel] (1,0.5) --
                node[midway,above,rightlabel] {$u_2$}
                (2,0.5);
            \draw[leftlabel] (2,0.5) -- 
                node[midway,above,leftlabel] {$a_1$}
                (3,0.5);
            \draw[rightlabel] (3,0.5) --
                node[midway,above,rightlabel] {$a_2$}
                (4,0.5);
          
        \end{scope}
        \draw[context] (6,-1.9) -- (7,-1.9);
        \node[context] at (6.5,-1.7) {$e$};
        \node[context] at (6.5,-2.5) {$\cdots$};
        \begin{scope}[yshift=-3cm,xshift=7cm]
            \draw (0,0) rectangle (6,1);
            \draw (2,0) -- (2,1);
            \draw (4,0) -- (4,1);
            \node at (1,-0.5) {$\aMgraph[U]$};
            \node at (3,-0.5) {$\aMgraph[A]$};
            \node at (5,-0.5) {$\aMgraph[V]$};
            \node at (1, 1.3) {$u$};
            \node at (3, 1.3) {$a$};
            \node at (5, 1.3) {$v$};
            \draw (0,1.1) -- (6,1.1);
            \draw[fill] (5,0.5) circle (0.05);
            \coordinate (NEWZ) at (5,0.5);

            \node at (5,0.2) {$z$};
            \draw[leftlabel] (4,0.5) --
                node[midway,above,leftlabel] {$v_1$}
                (5,0.5);
            \draw[rightlabel] (5,0.5) -- 
                node[midway,above,rightlabel] {$v_2$}
                (6,0.5);
        \end{scope}

        \draw[emb] (OLDX) to[bend right=30] (NEWX);
        \draw[emb] (OLDY) to[bend left=30] (NEWY);
        \draw[emb] (OLDZ) to[bend left=20] 
        node[midway, above] {$h$}
        (NEWZ);
    \end{tikzpicture}
    \caption{Illustration of the mapping $h$ defined in the proof 
        of \cref{bad-forest-paths-totally-ordered:lem}.}
    \label{splitting-idempotent:fig}
\end{figure}
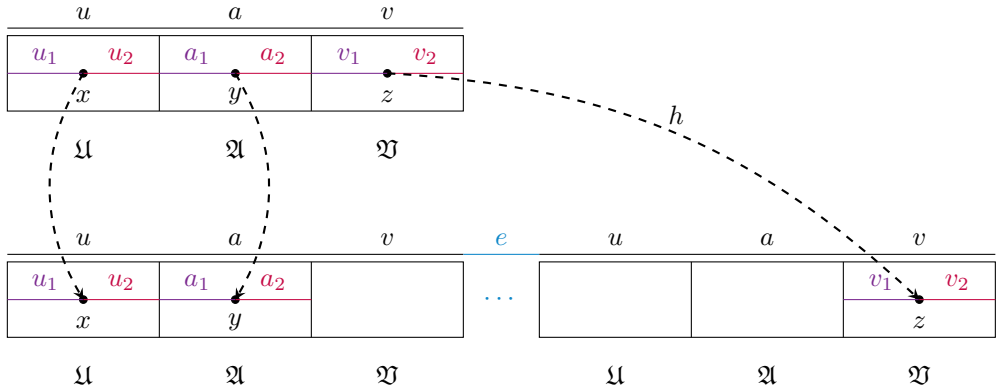

\begin{proofof}{bad-forest-paths-totally-ordered:lem}

    Let $e$ be an \kl{idempotent} element of $M$, and $\aMgraph$ be a
    \kl{monoid-labelled graph} that evaluates to $e$. By induction on the
    construction of $\aMgraph$, we can prove that there exists $u$,$a$,$v$ in
    $M$ satisfying $a \jequiv e$, and graphs $\aMgraph[U]$, $\aMgraph[A]$,
    $\aMgraph[V]$ such that they respectively evaluate to $u$, $a$, $v$,
    $\aMgraph[A]$ has a single vertex, and such that $\aMgraph = \aMgraph[U]
    \bp \aMgraph[A] \bp \aMgraph[V]$.

    Let us assume, without loss of generality that the label of the unique
    vertex of $\aMgraph[A]$ is a pair $(a_1, a_2)$ where $a_2 \jleq a_1$.
    We construct a function $h$ from vertices of
    $\aMgraph$ into vertices of
    $(\aMgraph[U] \bp \aMgraph[A] \bp \aMgraph[V])^3$, as follows:
    \begin{itemize}
        \item The function sends $\aMgraph[U]$ to the first copy of $\aMgraph[U]$ in the sequence
            acting as the identity,
        \item The function sends $\aMgraph[V]$ to the last copy of $\aMgraph[V]$ in the sequence
            acting as the identity,
        \item The function sends $\aMgraph[A]$ to the first copy of $\aMgraph[A]$ in the sequence
            acting as the identity.
    \end{itemize}

    We claim that the function $h$ defines an \kl(graph){embedding} of
    $\mcast{\aMgraph}$ into $\mcast{(\aMgraph[U] \bp \aMgraph[A] \bp
    \aMgraph[V])^3}$.  The graphical representation of this construction, where
    three distinguished vertices have been selected $x,y,z$ is given in
    \cref{splitting-idempotent:fig}. To prove that $h$ is an embedding, the
    only check to be made is that the monoid element \emph{between} $x$ and $y$
    (resp. $y$ and $z$, $x$ and $z$) is unchanged by the insertion of the new
    graphs in between. Let us do the proof in the case of $x$ and $z$, the
    remaining cases are similar.

    We have to prove that $u_2 a v_1 = u_2 a v e u a v_1$. Notice that because
    $e$ is idempotent, we know that $e^3 = u_1 u_2 a v e u a v_1 v_2 = u_1 u_2
    a v_1 v_2 = e$. Now, because $a_2 \jequiv e$, we conclude that $u_2 a v
    \jleq u_1$, hence using the cancellation property of
    \cref{cancellation:lem}, we can conclude that $u_2 a v e u a v_1 v_2 = u_2
    a v_1 v_2$. Similarly, we know that $u_2 a v_1 \jleq v_2$, and therefore we
    can use the cancellation property \cref{cancellation:lem} to conclude $u_2
    a v e u a v_1 = u_2 a v_1$. We have concluded.

    The other cases are similar, using the cancellation property whenever needed.

    To conclude we notice that if a path has length greater than $2$, one can split
    one inner graph using our construction.
\end{proofof}

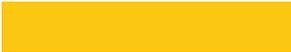
\end{document}